\documentclass[11pt, letterpaper]{article}

\usepackage{graphicx}
\usepackage{amsmath}
\usepackage{amssymb}
\usepackage{setspace}
\usepackage{epstopdf}
\usepackage{color}
\usepackage[letterpaper,left=1in,right=1in,top=1in,bottom=1in]{geometry}
\usepackage[]{algorithm2e}
\usepackage{multirow}
\usepackage{longtable}
\usepackage{rotating}
\usepackage{setspace}
\usepackage{geometry}
\usepackage{lineno}
\usepackage{float}
\usepackage{colortbl}
\usepackage{threeparttable}
\usepackage{arydshln}
\usepackage{mathtools}
\usepackage{amsthm}
\usepackage{booktabs}
\usepackage{makecell}
\usepackage[superscript]{cite}

\theoremstyle{plain}

\newtheorem{rmk}{Remark}

\newtheorem{alg}{Algorithm}

\title{\Large \bf Data-driven parallel Koopman subsystem modeling and distributed moving horizon state estimation for large-scale nonlinear processes}
\author{
\centerline{\normalsize Xiaojie Li$^{a}$, Song Bo$^{b}$, Xuewen Zhang$^{a}$, Yan Qin$^{c}$, Xunyuan Yin$^{a,}$\thanks{Corresponding author: X. Yin. Tel: +65-6316-8746. Email: xunyuan.yin@ntu.edu.sg}
}
\vspace{5mm}\\
\centerline{\small $^{a}$School of Chemistry, Chemical Engineering and Biotechnology, Nanyang Technological University}\\
\centerline{\small 62 Nanyang Drive, 637459, Singapore}\\
\centerline{\small $^{b}$Department of Chemical \& Materials Engineering, University of Alberta}\\
\centerline{\small Edmonton, AB T6G 1H9, Canada}\\
\centerline{\small $^{c}$Engineering Product Development Pillar, The Singapore University of Technology and Design}\\
\centerline{\small 8 Somapah Road, 487372, Singapore}
}
\allowdisplaybreaks

\begin{document}

\date{}
\maketitle

\setstretch{2}

\begin{abstract}
In this work, we consider a state estimation problem for large-scale nonlinear processes in the absence of first-principles process models. By exploiting process operation data, both process modeling and state estimation design are addressed within a distributed framework. By leveraging the Koopman operator concept, a parallel subsystem modeling approach is proposed to establish interactive linear subsystem process models in higher-dimensional subspaces, each of which correlates with the original nonlinear subspace of the corresponding process subsystem via a nonlinear mapping. The data-driven linear subsystem models can be used to collaboratively characterize and predict the dynamical behaviors of the entire nonlinear process. Based on the established subsystem models, local state estimators that can explicitly handle process operation constraints are designed using moving horizon estimation. The local estimators are integrated via information exchange to form a distributed estimation scheme, which provides estimates of the unmeasured/unmeasurable state variables of the original nonlinear process in a linear manner. The proposed framework is applied to a chemical process and an agro-hydrological process to illustrate its effectiveness and applicability. Good open-loop predictability of the linear subsystem models is confirmed, and accurate estimates of the process states are obtained without requiring a first-principles process model.
\end{abstract}

\section*{Introduction}

Large-scale and tightly integrated processes have been widely used in various industries. Structurally flexible and computationally efficient advanced control systems are needed to appropriately monitor and intervene in the operation of these processes to ensure operating safety, and to achieve efficient and sustainable process operation/production \cite{daoutidis2018integrating,CSML12CCE,Daoutidis2016184,S09JPC}.
Despite the promise, the development and implementation of advanced control systems for large-scale complex processes have been challenging. Specifically, due to the high nonlinearity of the process dynamics and rapid increase in the scale and the level of structural complexity, conventional decision-making paradigms are no longer favorable, as they can lead to difficulties in systems organization and maintenance, low fault tolerance of a decision-making system, and intractable online computation. The distributed framework is a very promising alternative for the monitoring and control of complex processes. A distributed decision-making system consists of local decision-making agents, each of which typically only needs to handle a sub-problem instead of the plant-wide problem for the entire process (known as a subsystem of the process). The local agents collaborate with each other based on real-time information exchange to assess the current process operating status and make informed decisions on the most appropriate control actions \cite{daoutidis2018integrating,CSML12CCE,Daoutidis2016184,chen2020machine,pourkargar2017impact,tang2021coordinating}. {\color{black}The distributed architecture holds the promise to be used for developing
flexible and scalable decision-making schemes for large-scale industrial systems.}
A complete decision-making system typically involves distributed state estimation and distributed control. Distributed estimation is used to provide real-time estimates of key process state variables that are not measured by the existing sensing instruments, and is an enabler for comprehensive process monitoring and informed decision-making of distributed control. In this work, we focus on the distributed state estimation problem for large-scale nonlinear processes, which has received less research attention as compared to the distributed (model predictive) control counterpart \cite{schneider2015convergence,farina2010moving,yin2018forming}.

In the existing literature, there have been some results on distributed state estimation for large-scale systems/processes. Distributed estimation algorithms based on deterministic state observers were proposed \cite{acikmese20141037,Hong2008846,yin2020distributed}. In existing literature \cite{KM08TSP,6166428,6075282,VD03TCST,Battistelli2016}, consensus-driven distributed Kalman filtering \cite{KM08TSP,6166428,6075282,VD03TCST} and extended Kalman filtering algorithms \cite{Battistelli2016} were proposed on the basis of sensor networks. To provide optimal estimates in the presence of constraints, distributed moving horizon estimation approaches were proposed \cite{yin2021consensus,FFS12IJRNC}.
More relevant results on distributed state estimation can be found in \cite{pourkargar2019distributed,ZL13JPC ,yin2018subsystem}.
It is worth mentioning that the aforementioned methods assume high-fidelity (first-principles) system/process models are available, and good estimation performance is premised on the high accuracy of the existing dynamic model. However, from an application perspective, it can be challenging to establish a first-principles model with accurate model parameters due to the availability of very limited first-principles knowledge -- this is especially critical when dealing with large-scale complex nonlinear processes.
In addition, for nonlinear processes, the use of a nonlinear dynamic model will complicate the design and implementation of local estimators. Therefore, to bridge the gap between theoretical research and practice in distributed estimation for complex processes, it is necessary to model the process dynamics accurately without requiring the availability of full first-principles knowledge and accurate values of the associated parameters.

One of the alternative solutions to describing the dynamics of a nonlinear process is to establish a data-driven model by leveraging the concept of the Koopman operator.  A Koopman operator is a linear representation of the dynamics of a nonlinear process that can be used to predict the future evolution of the process states. Specifically, by performing a nonlinear coordinate transformation, the temporal evolution of the nonlinear process states can be characterized by a higher-dimensional linear state-space model governed by the Koopman operator \cite{koopman1931hamiltonian,mezic2004comparison,9277915}.
However, in real applications, the exact Koopman operator associated with a nonlinear process may have infinite dimensions, and can be very difficult to find. This has motivated the exploration of a finite-dimensional approximation of the exact Koopman operator.
There have been algorithms that can be used to calculate an approximation of the exact Koopman operator by leveraging historical process data; for example, the Dynamic Mode Decomposition \cite{schmid2010dynamic} and its non-trivial extension -- Extended Dynamic Mode Decomposition \cite{williams2015data}, the Generalized Laplace Analysis \cite{mauroy2013isostables}, and the Galerkin Method \cite{froyland2014computational}.

In the context of process control, Koopman identification is an emerging enabler for developing data-driven decision-making and monitoring schemes for nonlinear processes/systems using linear control and estimation methods \cite{narasingam2019koopman,korda2018linear,son2022hybrid,narasingam2020application,kaiser2021data}.
Based on Koopman identification, a Lyapunov-based model predictive control (MPC) for nonlinear chemical processes was proposed in Narasingam and Kwon \cite{narasingam2019koopman}. In Korda and Mezi{\'c} \cite{korda2018linear}, a Koopman predictor was established as the basis of linear MPC for nonlinear systems.
{\color{black} In Son et al. \cite{son2022hybrid}, a hybrid Koopman MPC approach was proposed. In this work, multiple local Koopman models were developed to account for the local dynamics of a batch pulping process under different operating conditions.}
There have been fewer results on Koopman state estimation as compared to the control counterpart. A data-driven Koopman Kalman filtering was developed for power systems \cite{netto2018robust}. In Yin et al. \cite{yin2022c}, a data-driven moving horizon estimation algorithm was proposed based on an identified Koopman linear model for constrained nonlinear processes.
More results on Koopman-based control and estimation can be found in \cite{9277915,narasingam2020application,proctor2018generalizing,brunton2016koopman,peitz2020data,9304259,2023Koopman_MPC,2023Koopman_MPC_2,surana2017koopman}. Considering its merits in terms of being able to bypass first-principles modeling, the Koopman concept will be leveraged in this work for data-driven distributed state estimation.
However, the aforementioned methods focused on control and estimation within the centralized framework, and the associated Koopman-based identification methods are not favorable for developing distributed estimation schemes. The reasons are mainly threefold: 1)
 the identification of one single Koopman linear model that describes the dynamics of the entire process will become much more challenging as the scale and complexity of the processes increase; 2) the nonlinear coordinate transformation incurs a significant increase in the dimension of the state-space, and the number of observables of a single Koopman model for a complex process with many state variables can be excessively large; 3) as subsystem models are needed to design the local estimators within a distributed scheme, if a plant-wide model is to be used, then it needs to be partitioned into subsystem models appropriately before designing local estimators, which brings in additional difficulty and incurs more research efforts.
 For data-driven state estimation for large-scale processes within a distributed framework, it is indeed favorable to identify Koopman-based interactive subsystem models in parallel as the basis of the design of local estimators.

In this work, we address a data-driven distributed state estimation problem for large-scale constrained nonlinear processes by proposing a parallel subsystem modeling approach and a data-driven distributed moving horizon estimation algorithm.  Specifically, a data-driven parallel subsystem model identification method is proposed to establish subsystem models that interact with each other based on the concept of the Koopman operator, which collaboratively describes the dynamics of the entire process. Both subsystem states and inputs are projected onto a higher-dimensional space. In the proposed parallel subsystem modeling method, for each subsystem, the states of its interactive subsystems (of which the states have direct influences on the subsystem being considered) are treated as known inputs.
Based on the identified subsystem models, local estimators are designed as moving horizon estimators, each of which is capable of handling both the local dynamics and the interactive dynamics of the associated subsystem, in the presence of constraints on subsystem states, inputs, and disturbances. The proposed framework consisting of parallel subsystem identification and distributed moving horizon estimation is illustrated via applications to a chemical process and an argo-hydrological process.


\section*{Preliminaries}

\subsection*{Notation}

$\mathbb I$ is a set of consecutive positive integers $\mathbb I=\{1,\ldots,m\}$.
$\circ$ denotes the operator of the function composition.
$\text{col}\left(x_1,\ldots,x_m\right)$ is a column vector consisting of $x_1,x_2,\ldots,x_m$.
$\mathbb Z^*$ is a set of positive integers.
${\bf{I}}_{n}$ represents an identity matrix of dimension $n$. ${\boldsymbol{0}}_{a\times b}$ denotes an $a\times b$ zero matrix.
$\left\{x\right\}^{b}_{a}$ represents a column vector comprising a sequence of variable $x$ from time instant $a$ to instant $b$, i.e., $x(a), x(a+1),\ldots, x(b)$.
$\left\|x\right\|^2_P$ denotes the square of the weighted Euclidean norm of vector $x$, which is computed as $\left\|x\right\|^2_P = x^{\text{T}}P x$.
$\text{diag}\big\{{A_1, A_2\ldots }\big\} $ denotes a block diagonal matrix of which the blocks on the main diagonal are $A_1$, $A_2$, ...
${\hat x}(d|k)$ represents an estimate of $x(d)$ obtained at sampling instant $k$ ($k\geq d$).


\subsection*{Process description}

Let us consider a class of general nonlinear processes of medium- to large-scales that can be divided into $m$ subsystems; the dynamics of each subsystem $i$, $i\in\mathbb I$, are assumed to be with the following form:
\begin{subequations}\label{eq:subsystem}
\begin{align}
x_i(k+1) &= f_i\big(x_i(k), X_i(k), u_i(k) \big)\label{eq:subsystem:1}\\
y_i(k+1) &= h_i\big(x_i(k) \big)\label{eq:subsystem:2}
\end{align}
\end{subequations}
where $x_i\in {\mathbb X}_{i} \subset \mathbb R^{n_{x_i}} $ is the state vector of subsystem $i$; $u_i\in \mathbb U_i\subset \mathbb R^{n_{u_i}}$ is a vector of known inputs that affect the dynamics of subsystem $i$ directly, including manipulated inputs and/or known disturbances to the process; $X_i \in{\bar{\mathbb X}}_i \subset \mathbb R^{n_{X_i}}$ is a vector of the states of the subsystems that have direct interaction with subsystem $i$; $y_i \in \mathbb R^{n_{y_i}}$ is the sensor measurements vector of subsystem $i$; $f_i$ is a vector-valued nonlinear function characterizing the dependence of the future states of subsystem $i$ on the current states of the local subsystem and the interactive subsystems and inputs to the local subsystem; $h_i$ is the function of sensor measurements for subsystem $i$, $i\in\mathbb I$.  Note that set {$\bar{\mathbb X}_i$ can be determined based on $\mathbb X_j$, $\forall j\in\mathbb I_i$, where $\mathbb I_i$ ($i\in\mathbb I$) denotes a set of the indices of the subsystems, of which the states directly affect the dynamics of subsystem $i$. For an illustrating purpose, if the state of subsystem $i$ at the next sampling instant is dependent on the current states of subsystems $2$ and $5$ (i.e., $X_i$ in (\ref{eq:subsystem:1}) contain the states of subsystems $2$ and $5$), then $\mathbb I_i = \left\{2,5\right\}$.

It is worth mentioning that by incorporating all the subsystem states into an augmented vector, the dynamics of the entire process consisting of all the subsystems in (\ref{eq:subsystem}) can be described by the following general form:
\begin{equation}\label{eq:general:model}
\begin{aligned}
x(k+1)  = & ~f\big(x(k) ,u(k) \big)\\
 y(k)  = &~ h\big(x(k) \big)
\end{aligned}
\end{equation}
where $x=\text{col}\left(x_1,\ldots,x_m\right)\in \mathbb X \subset \mathbb R^{n_x}$ (with $n_x=\sum n_{x_i}$) represents the process state vector; $u$ is a vector of all the known inputs that are involved in $u_i$, $i\in\mathbb I$; $y=\text{col}\left(y_1,\ldots,y_m\right)$ is the vector of all the sensor measurements; $f =\text{col}\left(f_1, \ldots, f_m\right)$ is an augmented nonlinear vector field characterizing the dynamics of the entire process; $h =\text{col}\left(h_1, \ldots, h_m\right)$ describes the dependence of all the sensor measurements on the process state.

\subsection*{Problem formulation}
In this work, we aim to address the state estimation problem for the nonlinear process of medium- to large-scales in (\ref{eq:subsystem}) (or equivalently, (\ref{eq:general:model})) by proposing a distributed scheme consisting of local estimators developed based on subsystem models. However, as mentioned, it may not be favorable to utilize the subsystem models in (\ref{eq:subsystem}) directly for designing the local estimators due to: 1) the model structure and/or the values of the model parameters may not be completely known; 2) the development of local estimators based on nonlinear models are more challenging, and the implementation of nonlinear estimators is much more computationally complex than the case when linear models are used.

Based on the above observations, we aim first to construct data-driven linear subsystem models, which can be used to approximate the dynamics of the subsystems of the process in (\ref{eq:subsystem}) in a lifted function space that is related to the original function space via a nonlinear lifting mapping.
Specifically,
the model to be identified for subsystem $i$, $i\in\mathbb I$, is in the following form:
\begin{subequations}\label{eq:general:subsysmodel:Linear}
\begin{align}
z_i(k+1)  = & ~A_{ii}z_i(k) + \sum_{j\in\mathbb I_i} A_{ij} z_j(k) + B_i {\tilde u}_i(k)  \label{eq:general:subsysmodel:Linear:1}\\
y_i(k)  = &~ C_{i} z_i(k)  \label{eq:general:subsysmodel:Linear:2}\\
{\hat x}_i (k)  = &~ D_{i}z_i(k) \label{eq:general:subsysmodel:Linear:3}
\end{align}
\end{subequations}
where $z_i\in\mathbb R^{n_{z_i}}$ denotes the vector of lifted states of the linear model for subsystem $i$; ${\tilde u}_i\in\mathbb R^{n_{{\tilde u}_i}}$ denotes the vector of lifted inputs to the linear model for subsystem $i$; ${\hat x}_i\in\mathbb R^{n_{x_i}}$ denotes a prediction of the actual state of subsystem $i$ based on the evolution of the identified linear model (\ref{eq:general:subsysmodel:Linear:1}); $A_{ii}\in\mathbb R^{n_{z_i}\times n_{z_i}}$ and $A_{ij}\in\mathbb R^{n_{z_i}\times n_{z_j}}$ are subsystem matrices characterizing the dependence of future subsystem states on the local subsystem states and interactive subsystem states, respectively; $B_i\in\mathbb R^{n_{z_i} \times n_{{\tilde u}_i}}$ is an input matrix characterizing the dependence of the subsystem states on known inputs of subsystem $i$; $C_i\in\mathbb R^{n_{y_i}\times n_{z_i}}$ and $D_i\in\mathbb R^{n_{x_i}\times n_{z_i}}$ are output measurement and state-reconstruction matrices that need to be identified for subsystem $i$, $i\in\mathbb I$.

With the data-driven subsystem models in the form of (\ref{eq:general:subsysmodel:Linear}), local state estimators will be designed and a distributed estimation scheme will be formed for online estimating actual states of the considered nonlinear process.

\begin{rmk}
Deterministic models are presented in this section to facilitate the description of the proposed subsystem model identification method. It is worth noting that the parallel subsystem model identification and distributed moving horizon estimation algorithm proposed in this work can be used to handle unknown disturbances and measurement noise.
\end{rmk}

\section*{Parallel Koopman identification of the linear subsystem models}\label{section:3}
In this section, a parallel subsystem modeling method is proposed for establishing data-driven linear subsystem models in the form of (\ref{eq:general:subsysmodel:Linear}) as needed for the development of a distributed estimation scheme.

\subsection*{The basis of Koopman operator}\label{section:3:1}

First, the concept and the basis of the Koopman operator for discrete nonlinear controlled processes, which are leveraged for developing the subsystem modeling approach are reviewed. Specifically, consider the following nonlinear controlled system:
\begin{equation}\label{eq:general:model:uncontrol}
x(k+1) = F\big(x(k),u(k)\big)
\end{equation}
where $x\in\mathbb R^{n_x}$ is the vector of process states, $u\in\mathbb R^{n_u}$ is the vector of input vectors, and $F:\mathbb R^{n_x} \times \mathbb R^{n_u} \rightarrow \mathbb R^{n_x}$ is a vector-valued nonlinear function.

For system (\ref{eq:general:model:uncontrol}), a Koopman operator can be constructed
in an extended Euclidean space constituted by the Cartesian product of the space of the original process states $x$ and the space of the known inputs $u$. Specifically, an augmented vector containing both the system states and the known inputs is created as follows:
\begin{equation}
\mathcal X = \left[ \begin{array}{l}
x\\
u
\end{array} \right]
\end{equation}
which constitutes the state vector of the extended space. The dynamics of $\mathcal X$  be described by:
\begin{equation}\label{eq:augmented}
\mathcal X(k+1) = \mathcal F\left(\mathcal X(k)\right):= \left[ \begin{array}{c}
F\left(x(k),u(k)\right)\\
  \mathcal S u(k)
\end{array} \right]:= \left[ \begin{array}{c}
F\left(x(k),u(k)\right)\\
  u(k+1)
\end{array} \right]
\end{equation}
where $\mathcal S$ is a left shift operator, that is, $\mathcal S u(k) = u(k+1)$.

For the augmented system in (\ref{eq:augmented}), a linear dynamic model based on the Koopman operator concept is defined as \cite{korda2018linear,narasingam2020application,arbabi2018data}:
\begin{equation}\label{koopman:operator}
\mathcal K g =g \circ \mathcal F
\end{equation}
where $g$ represents the observables of a higher dimensional function space $\mathcal G$, which is associated with the original extended space (which is constituted by the Cartesian product of the process states $x$ and known inputs $u$) through a certain nonlinear mapping. $\mathcal K:\mathcal G\rightarrow \mathcal G$ is the Koopman operator governing the time evolution of the observables in the lifted space $\mathcal G$.
Specifically, for each sampling time $k\geq  0$, (\ref{koopman:operator}) indicates that:
\begin{equation}\label{koopman:operator2}
\mathcal K g\big(\mathcal X(k)\big) = g \circ \mathcal F\left(\mathcal X(k)\right) = g\big(\mathcal X(k+1)\big)
\end{equation}

One favorable property of the Koopman operator is its linearity despite the nonlinear dynamics of the original system, that is, for any pair of observables $g_1, g_2 \in \mathcal G$, and for any scalar coefficients $\alpha, \beta \in\mathbb R$, it is satisfied that:
\begin{equation*}
\begin{aligned}
\mathcal K\left(\alpha g_{1} + \beta g_{2}\right) &= \left(\alpha g_{1} + \beta g_{2}\right)\circ \mathcal F \\
&= \alpha g_{1}\circ \mathcal F +  \beta g_{2}\circ \mathcal F \\
&= \alpha \mathcal K g_{1} + \beta \mathcal K g_{2}
\end{aligned}
\end{equation*}
Consequently, the Koopman operator provides a linear representation of the time evolution of a nonlinear process in a higher-dimensional space, which is typical of infinite-dimension \cite{budivsic2012applied}.

\subsection*{EDMD-based parallel Koopman subsystem identification method}\label{section:3:3}

In this subsection, we present a Koopman-based parallel identification method that can be used to construct finite-dimensional linear subsystem models in the form of (\ref{eq:general:subsysmodel:Linear}) to approximate the dynamics of the subsystems of the entire large-scale nonlinear process.

To identify linear models for the $m$ subsystems instead of building a single Koopman linear model in the form of (\ref{koopman:operator2}) for the entire nonlinear process, one challenge is how to appropriately handle the subsystem interactions.  In the nonlinear subsystem models in (\ref{eq:subsystem}) that describe process dynamics, the dependence of the future state of subsystem $i$ on the current states of the interactive subsystems $ X_i(k)$ can be viewed as known inputs. Accordingly, in the $i$th data-driven subsystem model (\ref{eq:general:subsysmodel:Linear}) that needs to be constructed via identification, the vectors of lifted states of the interactive subsystems (i.e., $z_j$, $j\in\mathbb I_i$) can also be treated as known inputs that affect the future state of the local subsystem lifted state $z_i$. Considering the concept of Koopman operator described in the subsection titled ``The basis of Koopman operator", for the state vector $x_i$ of each subsystem $i$, $i\in\mathbb I$, in the form of (\ref{eq:subsystem}), there exists a (possibly infinite-dimensional) space $\mathcal G_i$, of which the observables can be used to describe the dynamics of the same subsystem using a Koopman operator in a linear state-space form as an analog of (\ref{koopman:operator2}), with both interactive subsystem states $X_i$ and subsystem inputs $u_i$ being treated as known inputs in Koopman identification. Specifically, construct an augmented state vector $\mathcal X_i$ being $\mathcal X_i:=\left[x_i^{\text{T}}, X_i^{\text{T}}, u_i^{\text{T}}\right]^{\text{T}}$, and its dynamics can be characterized by
\begin{equation}
\mathcal X_i(k+1) = \mathcal F_i\left(\mathcal X_i(k)\right):= \left[ \begin{array}{c}
f_i\left(x_i(k),X_i(k),u_i(k)\right)\\
  \mathcal R_i X_i(k)\\
  \mathcal S_i u_i(k)
\end{array} \right]= \left[ \begin{array}{c}
  x_i(k+1) \\
  X_i(k+1)\\
  u_i(k+1)
\end{array} \right]
\end{equation}
{\color{black}where $\mathcal R_i$ and $\mathcal S_i$ are left shift operators, that is, $\mathcal{R}_{i}X_{i}(k)=X_{i}(k+1)$ and $\mathcal{S}_{i}u_{i}(k)=u_{i}(k+1)$.}
This way, from a theoretical viewpoint, it is possible to use a Koopman operator to characterize the underlying local dynamical behavior of each nonlinear subsystem $i$ in the form of (\ref{eq:subsystem:1}) using a linear state-space model established via lifting the original extended space containing $\mathcal X_i$.

 However, the dimension of each space $\mathcal{G}_i$, $i\in\mathbb I$, and the corresponding exact Koopman operator are typically infinite. In addition, it can be very challenging to find exact Koopman operators for the subsystems for most nonlinear processes. These facts significantly limit the use of exact Koopman operators in the analysis of nonlinear processes using linear techniques. Instead of seeking the exact Koopman operator, we propose an alternative method to construct finite-dimensional approximated Koopman operators for the subsystems of the nonlinear process. In the existing literature, approximation algorithms that may be used to construct an approximated Koopman operator by leveraging batch process data include the Dynamic Mode Decomposition \cite{schmid2010dynamic} and its non-trivial extension -- Extended Dynamic Mode Decomposition (EDMD) \cite{williams2015data}, the Generalized Laplace Analysis \cite{mauroy2013isostables}, and the Galerkin Method \cite{froyland2014computational}.}  {\color{black}The EDMD algorithm is an effective approach to approximate Koopman operators in finite-dimensional state space. This approach utilizes a nonlinear mapping to transform the original state vectors into lifted states in a higher-dimensional space, where the dynamics are described by using a Koopman-based linear model. The EDMD method is capable of handling processes exhibiting strong nonlinearity \cite{williams2015data}, and is leveraged to develop the Koopman-based parallel subsystem identification method of the present work.}
%

%


Specifically, to establish a linear model for each subsystem $i$, $i\in\mathbb I$ in the form of (\ref{eq:general:subsysmodel:Linear}), the first step of the proposed method is to establish a nonlinear mapping from the original subsystem state vector $x_i$ to the subsystem lifted states on finite-dimensional space $\tilde {\mathcal G_i}$, which is a subspace of $\mathcal{G}_i$ and is spanned by a number of user-specified linearly independent state basis functions. On this subspace, $n_{z_i}$ state lifting functions, denoted by $\phi^i_{p}(x_i) :\mathbb X_i\rightarrow \mathbb R$, $\forall p=1,\ldots, n_{z_i}$, are defined for each subsystem $i$, $i\in\mathbb I$.
In addition, as an analog of the nonlinear mapping for each subsystem state vector $x_i$, a nonlinear mapping that transforms the original inputs $u_i$ into lifted inputs ${\tilde u}_i$ of the linear subsystem model (\ref{eq:general:subsysmodel:Linear}) is determined, and the corresponding input lifting functions are represented by $\delta^i_q (u_i):\mathbb U_i\rightarrow \mathbb R, q = 1,\ldots, n_{{\tilde u}_i}$ where $ n_{{\tilde u}_i}$ represents the dimension of the vector of lifted inputs $\tilde u_i$ to subsystem $i$.

For each subsystem $i$ in (\ref{eq:subsystem:1}), by leveraging the nonlinear mappings associated with the subsystems states and inputs, respectively, we define the vector of observables as follows:
\begin{equation}
\psi^i (\mathcal X_i)
= \left[
    \begin{array}{c}
        \phi^i(x_i) \\ \hdashline[2pt/2pt]
        \Phi^i(X_i) \\ \hdashline[2pt/2pt]
        \delta^i(u_i)
    \end{array}
\right]
\end{equation}
where $\phi^i$ is the vector of state lifting functions for subsystem $i$ (i.e., $\phi^i = \big[\phi^i_{1},\ldots,\phi^i_{n_{z_i}}\big]^{\text{T}}$); $\delta^i$ is the vector of input lifting functions for subsystem $i$ (i.e., $\delta^i = \big[\delta^i_{1},\ldots,\delta^i_{n_{{\tilde u}_i}}\big]^{\text{T}}$);  $\Phi^i$ is a concatenated vector consisting of state lifting functions of the interactive subsystems of subsystem $i$ (i.e., $\phi^j(x_j), \forall j\in\mathbb I_i$). This way, both the states of the interactive subsystems $X_i$ and the known inputs $u_{i}$, which are treated as external inputs to subsystem $i$ are taken into account in the observables.

Based on the defined vector of observables $\psi^i$, a finite-dimensional Koopman linear model can be established for each subsystem $i$, $i\in\mathbb I$, as follows:
\begin{equation}\label{koopman:operator:controlled:approx}
\psi^i \circ \mathcal F_i\left(\mathcal X_i(k)\right) = \psi^i \left(\mathcal X_i({k+1})\right)
\approx {\tilde{\mathcal K}}_i \psi^i \left(\mathcal X_i(k)\right)
\end{equation}
where 
 $\psi^i  \in \mathbb R^{n_i}$
($n_i \gg n_{x_i}+n_{X_i}+n_{u_i}$) is the subsystem observable vector;
$\tilde{\mathcal K_i}\in\mathbb R^{n_i\times n_i}$ is a finite-dimensional Koopman operator governing the time evolution of the subsystem observables in (\ref{koopman:operator:controlled:approx}).

The Koopman linear model (\ref{koopman:operator:controlled:approx}) will be utilized to establish a subsystem model for subsystem $i$ in the form of (\ref{eq:general:subsysmodel:Linear}), of which the lifted states are defined as $\phi^i$ and the lifted inputs are $\delta^i$; that is, $z_i = \phi^i$, with initial condition $z_i(0) =  \phi^i\left(x(0)\right)$, and ${\tilde u}_i=\delta^i$. This is based on the consideration that $\phi^i$ is only dependent on the state of the local subsystem, such that it may be easier to reconstruct the actual states $x_i$ of each subsystem $i$ in (\ref{eq:subsystem:1}) from the trajectories of the lifted states of (\ref{eq:general:subsysmodel:Linear:1}), which can be obtained based on the trajectories of the observables of each Koopman linear model (\ref{koopman:operator:controlled:approx}).

The next step is to calculate an approximated Koopman operator $\tilde{\mathcal K_i}$ for each subsystem $i$, $i\in\mathbb I$. It is desirable to make the approximation in (\ref{koopman:operator:controlled:approx}) as accurate as possible. Therefore, the objective is to find the Koopman operator that minimizes the discrepancy between the actual values of the observables and the predicted values of the observables of each subsystem. Considering the finite number of available data samples, for each subsystem $i$, $i\in\mathbb I$, we aim to find a $\mathcal{\tilde{K}}_i$ that provides the smallest averaged $L^2-$norm prediction error over the entire time window, which is described as follows:
\begin{equation}\label{eq:cost}
\mathcal{{\tilde K}}_i = \arg\mathop {\min }\limits_{\mathcal{{\tilde K}}_i } \sum_{k=1}^N \left\|\psi^i\left(\mathcal X_i(k+1)\right) - \mathcal{{\tilde K}}_i \psi^i\left(\mathcal X_i(k)\right)\right\|^2
\end{equation}
where $N$ denotes the total number of data samples available for analysis.

Considering the components of the subsystem observables, the approximated Koopman operator $\mathcal{{\tilde K}}_i$ can be expressed in the form of blocks as follows:
\begin{equation}\label{Koopman:block}
\mathcal{{\tilde K}}_i = \left[
    \begin{array}{c;{2pt/2pt}c;{2pt/2pt}c;{2pt/2pt}c;{2pt/2pt}c}
        A_{ii}&\ldots &A_{ij} & \ldots &B_i \\ \hdashline[2pt/2pt]
        *&* &\ldots &\ldots &* \\ \hdashline[2pt/2pt]
        \vdots&\vdots&  & &\vdots\\ \hdashline[2pt/2pt]
        *&*&\ldots &\ldots &*
    \end{array}
\right]
\end{equation}
In the first row of the matrix in (\ref{Koopman:block}), the elements between $A_{ii}$ and $B_i$ are constituted by blocks $A_{ij}$, $\forall j\in\mathbb I_i$, which are associated with $\Phi^i(X_i)$.
{\color{black}The primary goal of the proposed parallel subsystem modeling approach is to establish a linear subsystem model for each subsystem of the underlying nonlinear process, in the form of \eqref{eq:general:subsysmodel:Linear:1}. Each subsystem model characterizes the dynamic behaviors of its corresponding subsystem. To create a subsystem model for each subsystem $i$, $i\in\mathbb I$, we need to identify matrices $A_{ii}$, $A_{ij}$ for $j\in{\mathbb I}_i$, and $B_i$, which constitute the first-row blocks of the Koopman operator $\mathcal{{\tilde K}}_i$. The remaining blocks of $\mathcal{{\tilde K}}_i$ are not required for completing the subsystem model in the form of \eqref{eq:general:subsysmodel:Linear:1} for subsystem $i$. Therefore, only the blocks in the first row of the matrix on the right-hand side of (\ref{Koopman:block}) need to be identified, while all the remaining blocks are negligible.}
Consequently, for each subsystem $i$, $i\in\mathbb I$, the least-squares problem in (\ref{eq:cost}) is reformulated by retaining only the terms directly related to $\phi_i$ as follows:
\begin{equation}\label{eq:cost:AB}
\begin{aligned}
&A_{ii},\ldots, A_{ij}, \ldots, B_i \\
&= \arg\mathop {\min }\limits_{A_{ii},\ldots, A_{ij}, \ldots, B_i} \sum_{k=1}^N \bigg\|\phi^i\left(x_i(k+1)\right) - A_{ii} \phi^i\left(x_i(k)\right) - \sum_{j\in\mathbb I_i} A_{ij} \phi^j\left(x_j(k)\right) - B_i \delta^i\left(u_i(k)\right)\bigg\|^2\\[0.15em]
&= \arg\mathop {\min }\limits_{A_{ii},\ldots, A_{ij}, \ldots, B_i} \sum_{k=1}^N \bigg\|z_i(k+1) - A_{ii} \phi^iz_i(k) - \sum_{j\in\mathbb I_i} A_{ij} z_j(k) - B_i{\tilde u}_i(k)\bigg\|^2
\end{aligned}
\end{equation}

As an analog of (\ref{eq:cost:AB}), to identify the output measurement matrix $C_{i}$ in (\ref{eq:general:subsysmodel:Linear:2}) and the projection matrix $D_i$ in (\ref{eq:general:subsysmodel:Linear:3}) for subsystem $i$,  the following least-squares problems are formulated:
\begin{subequations}\label{eq:cost:CD}
\begin{align}
C_i =& \arg\mathop {\min }\limits_{C_i } \sum_{k=1}^N \left\|y_i(k)- C_i\phi^i( x_i(k))\right\|^2\label{eq:cost:CD:C}\\
D_i =& \arg\mathop {\min }\limits_{D_i } \sum_{k=1}^N \left\|x_i(k)- D_i\phi^i(x_i(k))\right\|^2\label{eq:cost:CD:D}
\end{align}
\end{subequations}

Since the formulated optimization problems in (\ref{eq:cost:AB}) and (\ref{eq:cost:CD}) are linear least-squares problems, they can be solved straightforwardly.
Next, we present an implementation algorithm describing the key steps that need to be followed to conduct paralleled subsystem model identification as follows:

\begin{alg}\label{Algorithm:1}
 Implementation algorithm for EDMD-based paralleled subsystem model identification
\begin{enumerate}
\item Collect $N+1$ ($N\gg n_i$) discrete samples for the process states $x$ and known inputs $u$;

\item For each subsystem $i$, $i\in\mathbb I$, do the following for data scaling:

\begin{enumerate}
\item [2.1.]  Scale each subsystem state variable $x_{i,l}$ (denoting the $l$th element of subsystem state vector $x_i$), $l=1,\ldots,n_{x_i}$ following $x_{i,l}(k) \leftarrow \frac{x_{i,l}(k)-x^{\min}_{i,l}}{x^{\max}_{i,l}-x^{\min}_{i,l}}$ where $x^{\max}_{i,l}:=\max\left\{x_{i,l}(k):k\in\mathbb Z^*\right\}$; $x^{\min}_{i,l}:=\min\left\{x_{i,l}(k):k\in\mathbb Z^*\right\}$.


\item [2.2.] Scale each subsystem input variable $u_{i,l}$ (the $l$th element of subsystem input vector $u_i$), and scale each sensor measurement $y_{i,l}$ (the $l$th element of vector $y_i$) in a way similar to Step 2.1.

\item [2.3.] Create subsystem snapshot matrices ${\boldsymbol {x_i}}$, ${\boldsymbol {{\bar x}_i}}$, ${\boldsymbol {u_i}}$ and ${\boldsymbol {y_i}}$ as: ${\boldsymbol {x_i}} = \left[x_i(1),\ldots,x_i(N)\right]$, ${\boldsymbol {{\bar x}_i}}  = \left[x_i(2),\ldots,x_i(N+1)\right]$, ${\boldsymbol {u_i}} = \left[u_i(1),\ldots,u_i(N)\right]$, and ${\boldsymbol {y_i}} = \left[y_i(1),\ldots,y_i(N)\right]$.

\item [2.4.] Specify a set of state lifting functions $\phi^i(x_i) = \big[\phi^i_{1}(x_i),\ldots,\phi^i_{n_{z_i}}(x_i)\big]^{\text{T}}$, with the first $n_{x_i}$ functions defined as $\phi^i_l(x_i) = x_{i,l}$, $l=1,\ldots,n_{x_i}$.

\item [2.5.] Specify a set of input lifting functions  $\delta^i(u_i) = \big[\delta^i_{1}(u_i),\ldots,\delta^i_{n_{{\tilde u}_i}}(u_i)\big]^{\text{T}}$.

\end{enumerate}

\item For each subsystem $i$, $i\in\mathbb I$, do the following in parallel for subsystem model identification:

\begin{enumerate}

\item [3.1] The identification agent for subsystem $i$ receives $\phi^j\left(\boldsymbol{x_i}\right)$ from the agents for each subsystem $j$, $j\in\mathbb I_i$.

\item [3.2] The identification agent for subsystem $i$ computes subsystem matrices $A_{ii}$, $A_{ij}(\forall j\in\mathbb I_i$), and $B_i$ following: $
 \left[\begin{array}{c;{2pt/2pt}c;{2pt/2pt}c}
       A_{ii}& \ldots,A_{ij},\ldots & B_i
    \end{array}\right]
 = {\phi^i}\left({\boldsymbol {{\bar x}_i}}\right) \cdot{\psi^i\left({\boldsymbol {\mathcal {X}_i}}\right)}^{\text{T}} \cdot\left(\psi^i\left({\boldsymbol {\mathcal {X}_i}}\right) \cdot {\psi^i\left({\boldsymbol {\mathcal X_i}}\right)}^{\text{T}}\right)^{+}$.

\item [3.3] The identification agent for subsystem $i$ computes output measurement matrix $C_i$ as: $C_i = {\boldsymbol {y_i}}\cdot
\phi^i (\boldsymbol {x_i})^{\text{T}} \cdot\left(
\phi^i (\boldsymbol {x_i}) \cdot
\phi^i (\boldsymbol {x_i})^{\text{T}}\right)^+$.
\end{enumerate}

\end{enumerate}
\end{alg}

In Step 3.2 of Algorithm~\ref{Algorithm:1},
$\psi^i \left({\boldsymbol {\mathcal X_i}}\right) = \left[\begin{array}{c;{2pt/2pt}c;{2pt/2pt}c}
       {\phi^i({\boldsymbol {x_i}})}^{\text{T}}&{\Phi^i({\boldsymbol {X_i}})}^{\text{T}} &{\delta^i({\boldsymbol {u_i}})}^{\text{T}}
    \end{array}\right]^{\text{T}} $.
In Step 2.4, for each subsystem $i$, $i\in\mathbb I$, the first $n_{x_i}$ state lifting functions are made identical to the states of the local subsystem, that is, $\phi^l_i(x_{i}) = x_{i,l}$, $l=1,\ldots,n_{x_i}$. This ensures the feasibility of reconstructing the original process states based on the prediction of the lifted state variables. Consequently, the  state-reconstruction matrix $D_i$ is determined directly as $D_i = \left[ {\begin{array}{*{20}{c;{2pt/2pt}c}}
{\bf{I}}_{n_{x_i}}& \bf{0}
\end{array}} \right]$, without needing to solve (\ref{eq:cost:CD:D}). Also, in many case scenarios, the measurement equations for the subsystems (\ref{eq:subsystem:2}) are linear, in the form of $y_i=H_i x_i$ (e.g., the sensor measurements are the temperatures of certain physical units belonging to one subsystem, which are also state variables of the subsystem). In such cases, matrix $C$ can be determined in a straightforward manner as $C_i = H_i D_i = \left[ {\begin{array}{*{20}{c;{2pt/2pt}c}}
H_i & \boldsymbol{0}_{n_{y_i}\times \left(n_{z_i}-n_{x_i}\right)}
\end{array}} \right]$ without resorting to Step~3.3.

\begin{rmk}
The optimal selection of the state basis functions $\phi^i$ and the input basis functions $\delta^i$ for each subsystem is still an open question for EDMD-based algorithms \cite{williams2015data,korda2018linear}. From a practical point of view, attempts can be made to find an appropriate set of basis functions that are ``rich enough" to account for the leading Koopman eigenfunctions, such that the Koopman-based subsystem models can be made sufficiently accurate. more detailed discussions pertaining to this point can be found in Williams et al. \cite{williams2015data}
\end{rmk}

\begin{rmk}
For many processes, the key variables have vastly different magnitudes (e.g., the mass fractions and temperatures which are usually considered process state variables); this may significantly affect the accuracy of the identified models.
By using Steps 2.2 and 2.3, the process state, input, and sensor measurement variables are scaled to be of similar magnitudes, such that almost equal importance is given to the variables and potential numerical issues can be avoided.
\end{rmk}

\section*{Partition-based distributed moving horizon estimation}\label{section:MHE}

In this section, we propose a linear partition-based distributed moving horizon estimation (MHE) method based on the linear subsystem models that are identified by applying the EDMD-based paralleled subsystem model identification approach proposed in the ``EDMD-based parallel Koopman subsystem identification method" subsection. {\color{black}An illustrative diagram that describes the connection between the subsystem modeling and the local estimator design of the proposed framework is shown in Figure~\ref{fig:diagram}.}
In addition, a schematic of the distributed estimation scheme is given in Figure~\ref{fig:DMHE:schematic}.
The distributed state estimation scheme consists of $m$ local MHE-based estimators interconnected with each other. Each local estimator accounts for the estimation of the states $z_i$ ($i\in\mathbb I$) of the corresponding subsystem in the higher-dimensional space. {\color{black} A communication network is deployed to account for the real-time information exchange among the different subsystems. At each sampling instant, each
MHE-based local estimator receives sensor measurements from its corresponding subsystem and communicates with the interacting subsystems to exchange subsystem state estimates and sensor measurements through the communication network.}

%

\subsection*{Subsystem models for the design of local estimators}
The identified Koopman subsystem models are used to develop the local MHE-based estimators of the distributed scheme in Figure~\ref{fig:DMHE:schematic}.
To account for unknown system disturbances and sensor measurement noise, a stochastic version of the identified model for subsystem $i$ is given as follows:
\begin{subequations}\label{eq:general:subsysmodel:Linear:stochas}
\begin{align}
z_i(k+1)  = & ~A_{ii}z_i(k) + \sum_{j\in\mathbb I_i} A_{ij} z_j(k) + B_i {\tilde u}_i(k) + w_i(k)  \label{eq:general:subsysmodel:Linear:stochas:1}\\
y_i(k)  = &~ C_{i} z_i(k) + v_i(k)  \label{eq:general:subsysmodel:Linear:stochas:2}\\
{\hat x}_i (k)  = &~ D_{i}z_i(k) \label{eq:general:subsysmodel:stochas:Linear:3}
\end{align}
\end{subequations}
where $w_i\in\mathbb W_i$ represents the disturbances and $v_i\in\mathbb V_i$ represents the measurement noise for subsystem $i$, with $\mathbb W_i \subset \mathbb R^{n_{z_i}}$ and $\mathbb V_i\subset \mathbb R^{x_{y_i}}$ being two compact sets.

\subsection*{Centralized MHE}

Before introducing the proposed distributed MHE design, we briefly review a representative centralized MHE design reported in Rao at al. \cite{rao2001constrained} Consider a general linear system model as follows:
\begin{equation}\label{eq:linear:model}
\begin{aligned}
z(k+1) &= A z(k) + B {\tilde u}(k) + w(k)\\
y(k) & = Cz(k) + v(k)
\end{aligned}
\end{equation}
where $z$ is the system state; $\tilde u$ is the known input vector; $w$ represents unknown disturbances; $y$ is the output measurements; $v$ is the measurement noise. Note that
(\ref{eq:linear:model}) maybe viewed as an aggregation of the identified linear subsystem models in (\ref{eq:general:subsysmodel:Linear:stochas}). For (\ref{eq:linear:model}), the centralized MHE design is presented as follows \cite{rao2001constrained}:
\begin{subequations}\label{eq:Koopman:mhe}
\begin{align}
&\mathop {\min }  \limits_{{ \hat z({k-N|k})},\left\{{ \hat w({d|k})}\right\}^{k-1}_{d=k-N}} \Theta\big(\hat z({k-N|k}),\hat w(k-N|k),\ldots, \hat w(k-1|k)\big) \label{eq:Koopman:mhe:cost}\\[0.2em]
&\textmd{\bf{ s.t.~}}
\quad {\hat z}({d+1|k})=A {\hat z}({d|k})+B{\tilde u}(d) + {\hat w}({d|k}),~~d=k-N,\ldots,k-1\label{eq:Koopman:mhe:model}\\[0.2em]
& ~~\quad\qquad\qquad y(d) = C{\hat z}({d|k}) + {\hat v}({d|k}),~~d=k-N,~\ldots,~k\label{eq:Koopman:mhe:v}
\end{align}
\end{subequations}
where the cost function is:
\begin{subequations}
  \begin{equation}\label{eq:cost:CMHE}
\Theta =
\sum\limits_{d = k-N}^{k - 1}  \left\|{\hat w}(d|k)\right\|^2_{Q^{-1}}+\sum\limits_{d = k-N}^{k} \left\|{\hat v}(d|k)\right\|^2_{R^{-1}}+\left\|{\hat z}(k-N|k)-{\bar z}(k-N)\right\|^2_{P_{k-N}^{-1}}
\end{equation}
{\color{black}The weighting matrix $P_{k}$ is updated using the covariance matrix update formula for the Kalman filter as:
\begin{equation}\label{eq:Pmatrix}
  P_{k} = Q+AP_{k-1}A^{\mathrm{T}}-AP_{k-1}C^{\mathrm{T}}(R+CP_{k-1}C^{\mathrm{T}})^{-1}CP_{k-1}A^{\mathrm{T}}
\end{equation}}
\end{subequations}
In (\ref{eq:Koopman:mhe}), $\hat z$ is an estimate of $z$; $\hat w$ and $\hat v$ are the estimates of $w$ and $v$, respectively; $Q$, $R$, and $P_{k}$ are weighting matrices; $N$ is the size of the estimation window; $\bar z(k-N)$ is a one-step ahead open-loop prediction of $z(k-N)$ computed following $\bar z (k-N) = A \hat z(k-N-1|k-1)+ B\tilde u(k-N-1)+\hat w(k-N-1|k-1)$.
At each sampling instant $k\geq N$, with the optimal estimates $\hat z({k-N|k})$ and $\left\{{ \hat w({d|k})}\right\}^{k-1}_{d=k-N}$ as well as model constraint (\ref{eq:Koopman:mhe:model}), the centralized MHE provides a sequence of the estimates $\hat z({k-N|k}),\ldots,\hat z({k|k})$ for the actual system states within the current estimation window. The last element of the sequence of the estimates (i.e., $\hat z({k|k})$) is used as the estimate of the state at the current instant.

\subsection*{Partition-based cost function for local MHE-based estimators}
In this subsection, we present the cost function for each local estimator of the proposed partition-based distributed MHE method, which is obtained based on partitioning the cost function for centralized MHE in (\ref{eq:Koopman:mhe}).

First, we define the individual cost function for each MHE-based estimator for subsystem $i$, $i\in\mathbb I$, of the distributed MHE by partitioning the cost function in (\ref{eq:cost:CMHE}).
Considering the state variables of each subsystem, $\Theta$ in (\ref{eq:cost:CMHE}) can be decomposed into $m$ partitioned functions $\Theta_i$, $i=1,\ldots,m$, such that the partitioned function associated with subsystem $i$ is with the following form:
\begin{subequations}
  \begin{equation}\label{eq:cost:DMHE:individual:1}
\Theta_i =
\sum\limits_{d = k-N}^{k - 1}  \left\|{\hat w_i}(d|k)\right\|^2_{Q_i^{-1}}+\sum\limits_{d= k-N}^{k} \left\|{\hat v_i}(d|k)\right\|^2_{R_i^{-1}}+\left\|{\hat z}_i({k-N|k})-{\bar z}_i(k-N)\right\|^2_{P_{i,k-N}^{-1}}
\end{equation}
where $P_{i,k}$ is updated using the covariance matrix update formula of distributed Kalman filter proposed in Li et al. \cite{li2023DEKF}
    \begin{align}\label{eq:Pmatrix:distributed}
    L_{i,k}&=(CA_{[:,i]}P_{i,k-1}A_{ii}^{\mathrm{T}}+C_{[:,i]}Q_{i})^{\mathrm{T}}(CA_{[:,i]}P_{i,k-1}A_{[:,i]}^{\mathrm{T}}C^{\mathrm{T}}+C_{[:,i]}Q_{i}C_{[:,i]}^{\mathrm{T}}+R)^{-1}\nonumber\\
P_{i,k}&=-L_{i,k}(CA_{[:,i]}P_{i,k-1}A_{ii}^{\mathrm{T}}+C_{[:,i]}Q_{i})+(A_{ii}P_{i,k-1}A_{ii}^{\mathrm{T}}+Q_{i})
  \end{align}
\end{subequations}
with $A_{[:,i]}$ and $C_{[:,i]}$ being matrices consisting of the columns of $A$ and $C$ that are associated with $z_{i}$, respectively.
{\color{black}By neglecting the coupling between the estimation errors of different subsystems, the summation of the arrival costs in (\ref{eq:cost:DMHE:individual:1}) for all the subsystems can be viewed as an approximation of the arrival cost in \eqref{eq:cost:CMHE} for the centralized MHE. }

{\color{black}Motivated by the designs of local objective functions for distributed MHE in
Schneider and Marquardt \cite{schneider2015convergence} and Li et al. \cite{li2023iterative} for each subsystem $i$, $i\in\mathbb I$, we incorporate sensor measurements from interacting subsystems into $\Theta_i$, which contains valuable information for state estimation of the local subsystem $i$, such that the cost function for each subsystem $i$, denoted by ${\tilde \Theta}_i$, is established as follows:
\begin{equation}\label{eq:cost:DMHE:individual:2}
\begin{aligned}
{\tilde \Theta}_i &=
\Theta_i  +\sum_{l\in\mathbb{I}_{i}}\Big\|\left\{{y_{l}}\right\}_{k-N}^k- \mathcal O^{l}_{[:,i]} {\hat z}_i(k-N|k)- \sum_{j\in\mathbb I_{i}}\mathcal O^{l}_{[:,j]} {\bar z}_j(k-N)
-\Lambda^{i} \left\{{\tilde u}\right\}_{k-N}^{k-1}  \nonumber\\
& \quad-\Gamma^{l}_{[:,i]} \left\{{\hat w}_i(d|k)\right\}_{d=k-N}^{k-1}\Big\|^2_{\boldsymbol{R}^{-1}}\\
&=\left\|{\hat z}_i({k-N|k})-{\bar z}_i(k-N)\right\|^2_{P_{i,k-N}^{-1}} +\sum\limits_{d = k-N}^{k - 1}  \left\|{\hat w_i}(d|k)\right\|^2_{Q_i^{-1}}+\bigg\|\
\left\{{ y}\right\}_{k-N}^k- \mathcal O_{[:,i]} {\hat z}_i(k-N|k)\nonumber\\
& \quad- \sum_{j\in\mathbb I_{i}}\mathcal O_{[:,j]} {\bar z}_j(k-N)
-\Lambda \left\{{\tilde u}\right\}_{k-N}^{k-1} -\Gamma_{[:,i]} \left\{{\hat w}_i(d|k)\right\}_{d=k-N}^{k-1}\bigg\|^2_{\boldsymbol{R}^{-1}}
\end{aligned}
\end{equation}
$\boldsymbol{R} := \text{diag}\big\{\underbrace{R,\ldots, R}_{N+1}\big\} $ with $R := \text{diag}\big\{R_1,\ldots, R_m\big\} $; $\Lambda,~\Gamma,~\mathcal O$ are with the following form:

\begin{equation*}
\begin{aligned}
{\Lambda} &:= \left[ {\begin{array}{*{20}{c}}
\boldsymbol{0}& \boldsymbol{0} & \cdots & \boldsymbol{0} & \boldsymbol{0}\\
CB&\boldsymbol{0}&{\cdots}&\boldsymbol{0}&\boldsymbol{0}\\
 \vdots &{\vdots}&{\cdots}&{\vdots}&{\vdots}\\
{C{A^{N - 2}}B}&{C{A^{N - 3}}B}&{\cdots}&\boldsymbol{0}&\boldsymbol{0}\\
{C{A^{N - 1}}B}&{C{A^{N - 2}}B}&{\cdots}&CAB&CB
\end{array}} \right];~~
{\Gamma} := \left[ {\begin{array}{*{20}{c}}
\boldsymbol{0}& \boldsymbol{0} & \cdots & \boldsymbol{0} & \boldsymbol{0}\\
C&\boldsymbol{0}&{\cdots}&\boldsymbol{0}&\boldsymbol{0}\\
 \vdots &{\vdots}&{\cdots}&{\vdots}&{\vdots}\\
{C{A^{N - 2}}}&{C{A^{N - 3}}}&{\cdots}&\boldsymbol{0}&\boldsymbol{0}\\
{C{A^{N - 1}}}&{C{A^{N - 2}}}&{\cdots}&CA&C
\end{array}} \right];\\
\mathcal O&:= \left[C^{\text{T}}~~\left(CA\right)^{\text{T}}~
\ldots~\left(CA^N\right)^{\text{T}}\right]^{\text{T}};\\
\end{aligned}
\end{equation*}
$\Gamma_{[:,i]}$ and $\mathcal O_{[:,i]}$ are matrices constituted of the columns of matrices $\Gamma$ and $\mathcal O$ that are associated with $w_i$ and $z_i$ of subsystem $i$, $i\in\mathbb I$, respectively; $\Gamma_{[:,i]}^{j}$ and $O_{[:,i]}^{j}$ denote the rows of the matrices $\Gamma_{[:,i]}$ and $O_{[:,i]}$ with respect to $z_{j}$, respectively; $\Lambda^{i}$ denotes the rows of the matrix $\Lambda$ that are associated with $z_{i}$.}

\subsection*{Formulation of the MHE-based estimators}
At each sampling instant $k\geq N$, Estimator $i$ (the local MHE-based estimator for the $i$th subsystem) will be used to provide the estimates of the subsystem states $\left\{{ \hat z_i({d|k})}\right\}^{k}_{d=k-N}$. Accordingly, the decision variables associated with Estimator $i$ include ${\hat z}_{i}(k-N|k)$ and $\left\{{ \hat w_i({d|k})}\right\}^{k-1}_{d=k-N}$, while ${\hat z}_j(k-N|k)$ and $\left\{{ \hat w_j({d|k})}\right\}^{k-1}_{d=k-N}$, $\forall j\in\mathbb I_{i}$ will be accounted for by the neighboring local estimators, and are treated as known constants to Estimator $i$ each time the optimization problem associated with Estimator $i$ is solved.

Consequently, by taking into account the hard constraints on the original process states $x$, the design of each estimator for subsystem $i$ is formulated as follows:

\noindent{\bf{\emph{Design of Local Estimators:}}}
At time instant $k\geq N$, given $\bar z_{i}(k-N)$, $\left\{{\bar y}\right\}_{k-N}^k$ , $\left\{{\tilde u}\right\}_{k-N}^{k-1}$, and ${\bar z}_{j}(k-N)$, $j\in\mathbb I_{i}$,  Estimator $i$, $i\in\mathbb I$, solves the following optimization problem:
\begin{subequations}\label{eq:cost:function:problem1:1}
\begin{align}
& { \hat z_i({k-N|k})},\big\{{ \hat w_i({d|k})}\big\}^{k-1}_{d=k-N} = \arg \mathop {\min }\limits_{{ \hat z_i({k-N|k})},\left\{{ \hat w_i({d|k})}\right\}^{k-1}_{d=k-N}}{\tilde \Theta}_{i}(k)\label{eq:cost:function:problem1:1:1}\\
&\textmd{\bf{ s.t.~}}  \qquad\qquad\quad~
\text{\bf{lb}}^i \leq {\boldsymbol{\hat Z}}^{(i)} \leq \text{\bf{ub}}^i\label{eq:cost:function:problem1:1:3}
\end{align}
\end{subequations}
In (\ref{eq:cost:function:problem1:1:1}), the cost function is:
\begin{equation}\label{eq:cost:function:theta:1}
\begin{aligned}
{\tilde \Theta}_{i}(k)
&= \left\|{\hat z}_{i}({k-N|k})-{\bar z}_i(k-N)\right\|^2_{P_{i, k-N}^{-1}} +\sum\limits_{d = k-N}^{k - 1}  \left\|{\hat w_i}(d|k)\right\|^2_{Q_i^{-1}}+\bigg\|\
\left\{{ y}\right\}_{k-N}^k- \mathcal O_{[:,i]} {\hat z}_i(k-N|k)\\
&~~~~ - \sum_{j\in\mathbb I_{i}}\mathcal O_{[:,j]} {\bar z}_j(k-N)
-\Lambda \left\{{\tilde u}\right\}_{k-N}^{k-1} -\Gamma_{[:,i]} \left\{{\hat w}_i(d|k)\right\}_{d=k-N}^{k-1}\|^2_{\boldsymbol{R}^{-1}}
\end{aligned}
\end{equation}
wherein the weighting matrix $P_{i,k}$ is updated following \eqref{eq:Pmatrix:distributed};
${\bar z}_{i}(k-N)$ is an {\emph {a priori}} prediction of $z_i(k-N)$, which is computed using the subsystem model as:
\begin{equation}\label{posteriori:eq}
{\bar z}_{i}(k-N) =
 A_{ii} {\hat z}_i(k-N-1|k-1) + \sum_{j\in \mathbb I_{i}}  A_{ij} {\hat z}_j(k-N-1|k-1)+B_i {\tilde u}_i(k-N-1) +\hat w_i(k-N-1|k-1)
\end{equation}
where ${\hat z}_i(k-N-1|k-1)$ and $\hat w_i(k-N-1|k-1)$, respectively, denote the estimates of $z_i(k-N-1)$ and $ w_i(k-N-1)$ provided by Estimator $i$  at time instant $k-1$; $N$ is the length of the estimation window of the local estimator.

Equation (\ref{eq:cost:function:problem1:1:3}) is an inequality constraint imposed on the decision variables (i.e., ${ \hat z_i({k-N|k})}$ and $\big\{{ \hat w_i({d|k})}\big\}^{k-1}_{d=k-N}$) such that the estimates of the lifted states of each subsystem will be constrained within the desired range determined by the constraints on the original process states $x$. Specifically,
${\boldsymbol{\hat Z}}^{(i)} :=\mathcal G_{[:,i]} {\hat z}_i(k-N|k)+ \sum_{j\in\mathbb I_{i}}\mathcal G_{[:,j]} {\bar z}_j(k-N)
+\mathcal H \left\{{\tilde u}\right\}_{k-N}^{k-1} + {\mathcal J}_{[:,i]} \left\{{\hat w}_i(d|k)\right\}_{d=k-N}^{k-1}$, in which:
\begin{equation*}
\begin{aligned}
{\mathcal H} &:= \left[ {\begin{array}{*{20}{c}}
\boldsymbol{0}& \boldsymbol{0} & \cdots & \boldsymbol{0} & \boldsymbol{0}\\
B&\boldsymbol{0}&{\cdots}&\boldsymbol{0}&\boldsymbol{0}\\
 \vdots &{\vdots}&{\cdots}&{\vdots}&{\vdots}\\
{{A^{N - 2}}B}&{{A^{N - 3}}B}&{\cdots}&\boldsymbol{0}&\boldsymbol{0}\\
{{A^{N - 1}}B}&{{A^{N - 2}}B}&{\cdots}&AB&B
\end{array}} \right],~~
{\mathcal J} := \left[ {\begin{array}{*{20}{c}}
\boldsymbol{0}& \boldsymbol{0} & \cdots & \boldsymbol{0} & \boldsymbol{0}\\
&\boldsymbol{0}&{\cdots}&\boldsymbol{0}&\boldsymbol{0}\\
 \vdots &{\vdots}&{\cdots}&{\vdots}&{\vdots}\\
{{A^{N - 2}}}&{{A^{N - 3}}}&{\cdots}&\boldsymbol{0}&\boldsymbol{0}\\
{{A^{N - 1}}}&{{A^{N - 2}}}&{\cdots}&A&I
\end{array}} \right],\\
\mathcal G&:= \left[I~~\left(A\right)^{\text{T}}~
\ldots~\left(A^N\right)^{\text{T}}\right]^{\text{T}},\\
\end{aligned}
\end{equation*}
${\mathcal J}_{[:,i]}$ and $\mathcal G_{[:,i]}$ are matrices composed of the columns of ${\mathcal J}$ and $\mathcal G$ that are associated with $w_i$ and $z_i$, $i\in\mathbb I$, respectively.
${\boldsymbol{\hat Z}}^{(i)}$ are composed of the estimates of states of the $i$th subsystems within the current estimation window, that is,
${\boldsymbol{\hat Z}}^{(i)} = \big[{\hat z}_i(k-N|k)^{\text{T}},~\ldots,~{\hat z}_i(k|k)^{\text{T}}\big]$. 
Meanwhile, for Estimator $i$, we only need to impose constraints on ${\hat z}_i$, as the estimates of the other subsystems are handled by the neighboring estimators. Accordingly,
$\text{\bf{lb}}^i$ and $\text{\bf{ub}}^i$ in (\ref{eq:cost:function:problem1:1:3}), which are the lower and upper bounds on ${\boldsymbol{\hat Z}}^{(i)}$, are determined in a way such that only constraints are imposed on the elements corresponding to the first $n_{x_i}$ states of ${\hat z}_i$, that is, ${\hat z}_i{({d|k})_{[1:n_{x_i}]}}$, for $d=k-N,\ldots,k$.
The remaining elements in $\text{\bf{lb}}^i$ and $\text{\bf{ub}}^i$ that are corresponding to each subsystem $j$, $j\in\mathbb I_i$, are set to be {\color{black}$-\infty$ and $+\infty$, respectively.}

By incorporating constraint (\ref{eq:cost:function:problem1:1:3}) with the defined $\text{\bf{lb}}^i$ and $\text{\bf{ub}}^i$ into each local MHE-based estimator, it is ensured that ${\hat z}_i({{d|k})_{[1:n_{x_i}]}}\in\mathbb X_i$, for $d=k-N,\ldots,k$.
Since these $n_{x_i}$ elements of the lifted state vector $z_i$ are made identical to $x_i$ according to Algorithm~\ref{Algorithm:1}), this constraint ensures that the estimates of the original states of each subsystem $x_i$ satisfy the physical hard constraints on them.

Note that the estimator design is presented without considering the scaling of the data samples.
Since the data used in parallel subsystem identification is scaled in implementation, the estimates provided by the local estimators of the distributed MHE scheme do not account for the actual process states. Therefore, the determination of the bounds $\text{\bf{lb}}^i$ and $\text{\bf{ub}}^i$ based on the constraints on the original subsystem state $x_i$ is also subject to scaling. In addition, the state estimates provided by the MHE-based estimators for the subsystems need to be unscaled, and the unscaled values will be used to represent the estimates of the actual process states.

\subsection*{Implementation algorithm}

We summarize the implementation steps for the proposed distributed estimation method in an algorithm as follows:

\begin{alg} \label{alg:2}
Implementation procedure for the distributed MHE scheme. \vspace{2mm}

At time ${k \geq N+1}$, do the following steps:

    \begin{enumerate}

      \item [1.] Estimator $i$, $i\in\mathbb I$, receives sensor measurements from all the subsystems $y(k)$.

      \item [2.] Estimator $i$, $i\in\mathbb I$, computes ${\bar z}_{i}(k-N)$ following  (\ref{posteriori:eq}).

            \begin{itemize}

              \item [2.1] Estimator $i$, $i\in\mathbb I$, requests and receives $\bar z_j(k-N|k)$ (i.e., the prediction of the lifted states of subsystem $j$), from the each Estimator $j$, $j\in\mathbb I_{i}$.

                \item [2.2.] Estimator $i$, $i\in\mathbb I$, solves (\ref{eq:cost:function:problem1:1}) to provide optimal estimates of the subsystem state and disturbances sequence, i.e., ${ \hat z_i({k-N|k})}$ and $\big\{{ \hat w_i({d|k})}\big\}^{k-1}_{d=k-N}$.

              \end{itemize}

        \item [3.] A sequence of optimal state estimates $\hat z({d|k})$, $d=k-N\ldots,k$, is generated for the state vector $z$ of entire process in the lifted space.
        \item [4.] Update the weighting matrix $P_{i,k}$ following \eqref{eq:Pmatrix:distributed}. Set $k=k+1$, go to Step 1.
    \end{enumerate}

\end{alg}


\begin{rmk}
With convex sets that bound the estimates of the subsystem states and disturbances, the execution of the local estimators can be implemented as a convex quadratic problem instead of solving non-convex and computationally expensive optimization problems associated with nonlinear MHE. This way, as compared to the case when estimators are developed based on first-principles nonlinear models, the computational efficiency of the proposed distributed estimation scheme is also expected to be improved despite the dimensionality increase of the subsystem models.
\end{rmk}

{\color{black}

\section*{Application to a chemical process}
In this section, the proposed framework is illustrated via a simulated application to a chemical process consisting of four continuous stirred tank reactors.
\subsection*{Process description}
This process involves four continuous stirred tank reactors (CSTRs) that are interconnected via mass and energy flows. A schematic of the process is presented in Figure \ref{cstr}.
A feed stream, containing pure reactant $A$, enters the $i$th reactor with flow rate $F_{0i}$, temperature $T_{0i}$, and molar concentration $C_{A0i}$. The effluent of the first reactor is fed into the second reactor at flow rate $F_{1}$, temperature $T_{1}$, and molar concentration $C_{A1}$. A fraction of the outlet stream from the second reactor is sent to the third reactor, with flow rate $F_{2}$, temperature $T_{2}$, and concentration  $C_{A2}$. The remaining portion of the effluent of the second reactor is sent back to the first reactor at flow rate $F_{r1}$, temperature $T_{2}$, and concentration $C_{A2}$. The outlet of the third reactor enters the fourth reactor at flow rate $F_{3}$, temperature $T_{3}$, and molar concentration $C_{A3}$. A portion of the outlet stream from the fourth reactor is recycled back to the first reactor at flow rate $F_{r2}$, temperature $T_{4}$, and concentration $C_{A4}$. Each of the four reactors is equipped with a jacket, which enables the addition or extraction of heat to or from its respective reactor.

Based on mass and energy balances, a first-principles dynamic model that consists of eight differential equations is presented as follows \cite{rashedi2017triggered}:
\begin{subequations}\label{eq:cstr}
\begin{align}
  \frac{\mathrm{d}T_{1}}{\mathrm{d}t} & = \frac{F_{01}}{V_{1}}(T_{01}-T_{1})+\frac{F_{r1}}{V_{1}}(T_{2}-T_{1})+\frac{F_{r2}}{V_{1}}(T_{4}-T_{1})-\sum_{i=1}^{3}\frac{\Delta H_{i}}{\rho c_{p}}k_{i0}e^{\frac{-E_{i}}{RT_{1}}}C_{A1}+\frac{Q_{1}}{\rho c_{p}V_{1}}\\
    \frac{\mathrm{d}C_{A1}}{\mathrm{d}t}&=  \frac{F_{01}}{V_{1}}(C_{A01}-C_{A1})+\frac{F_{r1}}{V_{1}}(C_{A2}-C_{A1})+\frac{F_{r2}}{V_{1}}(C_{A4}-C_{A1})-\sum_{i=1}^{3}k_{i0}e^{\frac{-E_{i}}{RT_{1}}}C_{A1}\\
     \frac{\mathrm{d}T_{2}}{\mathrm{d}t} & =\frac{F_{1}}{V_{2}}(T_{1}-T_{2})+\frac{F_{02}}{V_{2}}(T_{02}-T_{2})-\sum_{i=1}^{3}\frac{\Delta H_{i}}{\rho c_{p}}k_{i0}e^{\frac{-E_{i}}{RT_{2}}}C_{A2}+\frac{Q_{2}}{\rho c_{p}V_{2}} \\
    \frac{\mathrm{d}C_{A2}}{\mathrm{d}t}& = \frac{F_{1}}{V_{2}}(C_{A1}-C_{A2})+\frac{F_{02}}{V_{2}}(C_{A02}-C_{A2})-\sum_{i=1}^{3}k_{i0}e^{\frac{-E_{i}}{RT_{2}}}C_{A2}\\
     \frac{\mathrm{d}T_{3}}{\mathrm{d}t} &=  \frac{F_{2}-F_{r1}}{V_{3}}(T_{2}-T_{3})+\frac{F_{03}}{V_{3}}(T_{03}-T_{3})-\sum_{i=1}^{3}\frac{\Delta H_{i}}{\rho c_{p}}k_{i0}e^{\frac{-E_{i}}{RT_{3}}}C_{A3}+\frac{Q_{3}}{\rho c_{p}V_{3}}\\
    \frac{\mathrm{d}C_{A3}}{\mathrm{d}t}& = \frac{F_{2}-F_{r1}}{V_{3}}(C_{A2}-C_{A3})+\frac{F_{03}}{V_{3}}(C_{A03}-C_{A3})-\sum_{i=1}^{3}k_{i0}e^{\frac{-E_{i}}{RT_{3}}}C_{A3}\\
     \frac{\mathrm{d}T_{4}}{\mathrm{d}t} &=  \frac{F_{3}}{V_{4}}(T_{3}-T_{4})+\frac{F_{04}}{V_{4}}(T_{04}-T_{4})-\sum_{i=1}^{3}\frac{\Delta H_{i}}{\rho c_{p}}k_{i0}e^{\frac{-E_{i}}{RT_{4}}}C_{A4}+\frac{Q_{4}}{\rho c_{p}V_{4}}\\
    \frac{\mathrm{d}C_{A4}}{\mathrm{d}t}&  = \frac{F_{3}}{V_{4}}(C_{A3}-C_{A4})+\frac{F_{04}}{V_{4}}(C_{A04}-C_{A4})-\sum_{i=1}^{3}k_{i0}e^{\frac{-E_{i}}{RT_{4}}}C_{A4}
\end{align}
\end{subequations}
where $C_{Ai}$ represents the molar concentration of $A$ in $i$th reactor, $i=1,2,3,4$; $T_{i}$ is the temperature of feed stream in $i$th reactor, $i=1,2,3,4$.
The definitions of the remaining variables and parameters, and the values of the parameters involved in model \eqref{eq:cstr} can be found in Yin and Liu \cite{yin2020distributed} and Rashedi et al.\cite{rashedi2017triggered}

%
\subsection*{Problem formulation and simulation setting}
In this process, the temperatures in the four reactors can be measured using sensors, while the concentrations of the reactant $A$ need to be estimated online.
To achieve this objective, the process is decomposed into four subsystems according to the physical topology of the process, that is, each reactor is one subsystem. Then, the proposed method is applied: 1) to build Koopman subsystem models for the four subsystems to collaboratively characterize the dynamics of the entire process; 2) to establish a data-driven distributed state estimation scheme for estimating the full-state of the process.
In light of the objective, in this work, the nonlinear model in \eqref{eq:cstr} is not used for designing state estimators. Instead, it is only leveraged to generate data which will be used to establish data-driven Koopman subsystem models based on the proposed method.

The values of the model parameters in \eqref{eq:cstr} are made the same as those in Yin and Liu \cite{yin2020distributed}. The initial condition of the simulation for generating state trajectories is $[326.3794~\mathrm{K}~~3.1833~\mathrm{kmol/m^{3}}$ $326.3745~\mathrm{K}~~2.9402~\mathrm{kmol/m^{3}}~~328.0896~\mathrm{K}~~2.9863~\mathrm{kmol/m^{3}}~~326.7154~\mathrm{K}~~3.1649~\mathrm{kmol/m^{3}}]^{\mathrm{T}}$.
The sequence of heating input $Q_{i}$, $i=1,2,3,4$, is generated following a uniform distribution within predefined bounds, and it varies every 1.5 hours. Specifically, the upper bounds on the heating inputs to the four reactors are determined as
$Q_{1, \max}=1.2\times 10^{4} ~\mathrm{kJ/h}$, $Q_{2, \max}=2.2\times 10^{4}~ \mathrm{kJ/h}$, $Q_{3, \max}=2.7\times 10^{4}~ \mathrm{kJ/h}$, and $Q_{4, \max}=1.2\times 10^{4} ~\mathrm{kJ/h}$, while the lower bounds of the heating inputs are set as $Q_{1, \min}=0.8\times 10^{4}~\mathrm{kJ/h}$, $Q_{2, \min}=1.8\times 10^{4}~ \mathrm{kJ/h}$, $Q_{3, \min}=2.3\times 10^{4} ~ \mathrm{kJ/h}$, and $Q_{4, \min}=0.8\times 10^{4}~ \mathrm{kJ/h}$.
Unknown system disturbances are generated following Gaussian distribution with zero mean and standard deviation of $\theta_{w}$ and are then made bounded within $[-\bar{w}, \bar{w}]$, where $\bar{w}=5\theta_{w}$ with $\theta_{w}=[0.1554~\mathrm{K} ~~0.0015~\mathrm{kmol/m^{3}}~~0.1554~\mathrm{K}~~0.0014~\mathrm{kmol/m^{3}}~~0.1562~\mathrm{K}~~0.0014~\mathrm{kmol/m^{3}}~~0.1556~\mathrm{K}$ $0.0015~\mathrm{kmol/m^{3}}]^{\mathrm{T}}$. Similarly, unknown measurement noise is generated following Gaussian distribution with zero mean and standard deviation of $\theta_{v}$ and is then made bounded within $[-\bar{v}, \bar{v}]$, where $\bar{v}=5\theta_{v}$ with $\theta_{v}=[0.3108~\mathrm{K} ~~0.3108~\mathrm{K}~~0.3125~\mathrm{K}~~0.3112~\mathrm{K}]^{\mathrm{T}}$. The sampling period is selected as 0.025 \textrm{h}.
2000 samples within an operating period of 50 hours are recorded.
The 2000 samples are divided into three datasets.
The first dataset contains the first 1000 samples within the first 25-hour period, which will be used for Koopman subsystem modeling. The remaining 1000 samples are divided evenly into two groups: 500 samples, which constitute the second dataset, will be used to validate the data-driven subsystem models; the other 500 samples constitute the third dataset, which will be used to verify the performance of the proposed data-driven distributed state estimation scheme.

%


In this subsection, we apply the parallel subsystem modeling approach proposed in the ``EDMD-based parallel Koopman subsystem identification method" subsection to establish four Koopman models for the subsystems of the chemical process.
Given that the states assigned to different subsystems have the same physical meaning, we utilize the same state and input lifting functions $\phi^i$ and $\delta^i$, $i=1,2,3,4$, for different subsystems. Specifically, the lifting functions are chosen as the original subsystem state $x_i$, power function $\sqrt[3]{x_{i}}$, and exponential function $e^{x_{i}}$, that is,
\begin{equation*}
\phi^i =\left[ {\begin{array}{*{20}{c}}
T_{i}&C_{Ai}&\sqrt[3]{T_{i}}&\sqrt[3]{C_{Ai}}&{e^{T_{i}}}&{e^{C_{Ai}}}
\end{array}} \right]^{\text{T}},~\forall i=1,2,3,4.
\end{equation*}
The input lifting functions for each subsystem are selected as
$\delta^{i}(Q_i) =\left[\begin{array}{cc}
                          Q_{i} & \sqrt[3]{Q_{i}}
                        \end{array}\right]^{\text{T}},~\forall i=1,2,3,4.$
\subsection*{Koopman subsystem modeling}

The 1000 samples in the first dataset are scaled following Steps~2.1 and 2.2 of Algorithm~\ref{Algorithm:1}, and then are utilized for identifying the Koopman-based subsystem models.
Based on the selected state and input lifting functions, four subsystem models in the form of \eqref{eq:general:subsysmodel:Linear:stochas} are established by executing Algorithm~\ref{Algorithm:1}.
For each subsystem model, the lifted state vector $z_i$ has 6 variables; the lifted input vector $\tilde u_{i}$ for the $i$th subsystem, $i=1,2,3,4$, has 2 variables.

Cross-validation is conducted based on the second dataset which has 500 samples.
The trajectories of the actual states and the open-loop predictions of the states provided by the four Koopman-based subsystem models are presented in
Figure \ref{fig:cstr_validation}. Without incorporating state measurements into generating the state predictions, the Koopman subsystem models are able to provide accurate predictions for the temporal trajectories of all eight state variables.

\subsection*{Distributed state estimation results}
Based on the established Koopman subsystem models, a distributed state estimation scheme consisting of four MHE-based local estimators is developed.
Each local estimator is designed following \eqref{eq:cost:function:problem1:1}.
The length of the estimation horizon for each local estimator is $N=3$. The weighting matrices $P_{i,0}$, $Q_{i}$, and $R_{i}$ are selected as
  $P_{i,0}=\mathrm{diag}\{0.01,0.01,0.01,0.01, 0.01, 0.01\}$, $Q_{i}=\mathrm{diag}\{0.1,0.1,0.1,0.1, 0.1, 0.1\}$, and $R_{i}=\mathrm{diag}\{0.001,0.001, 0.001\}$,
for $i=1,2,3, 4$. It is noted that the estimators provide estimates of the scaled states, which are further unscaled to reconstruct the original process states.

The initial state of the process and the initial guess values adopted by the local estimators are presented in Table \ref{tbl:initial_states}. Hard constraints are incorporated into the local estimators of the distributed scheme, such that the estimates of the molar concentrations of the reactants are ensured to be non-negative.
The trajectories of the actual states and estimates of the full-state of the process are shown in Figure~\ref{fig:cstr_estimation}. The results confirm that the developed data-driven distributed estimation scheme can provide accurate estimates of the eight process states in the presence of disturbances and measurement noise.
%

To demonstrate the superiority of the proposed method, we compare the proposed estimation design with a
distributed MHE scheme where the local estimators are developed based on the linearized subsystem models obtained via Taylor series expansions. {\color{black}Specifically, we consider a steady-state operating point $x_{s}=[310.8376~\mathrm{K}~~3.0317~\mathrm{kmol/m^{3}}$ $310.8329~\mathrm{K}~~2.8002~\mathrm{kmol/m^{3}}$ $312.4663~\mathrm{K}~~2.844~\mathrm{kmol/m^{3}}$ $311.1576\,\mathrm{K}~3.0142~\mathrm{kmol/m^{3}}]^{\mathrm{T}}$, which corresponds to constant inputs $Q_{1}=1.0\times10^{4}~\mathrm{kJ/h}$, $Q_{2}=2.0\times10^{4}~\mathrm{kJ/h}$, $Q_{3}=2.5\times10^{4}~\mathrm{kJ/h}$, and $Q_{4}=1.0\times10^{4}~\mathrm{kJ/h}$. At $x_{s}$, a linearized model is obtained using Taylor-series expansion. The linearized model is further discretized with a step size $h=1.5~\mathrm{min}$, and a discrete-time model of the entire system is obtained. Then, the linearized global model is partitioned into linearized subsystem models in the form of \eqref{eq:general:subsysmodel:Linear} for the subsystems of the entire process.}
The root mean square error (RMSE) and the average computation time for the one-step execution of one local estimator for both designs are presented in Table \ref{tb2}.
Compared to the distributed estimation scheme based on the linearized subsystem models, the estimation error of the proposed method is significantly smaller. This is attributed to the much better predictability of the Koopman subsystem models when the operation is distant from the steady-state point at which the linearized subsystem models are obtained.
It is mentioned that the RMSE is computed based on the scaled process states and the corresponding estimates, such that equitable importance is given to the states with different physical meanings.
The average computation time for both designs, as presented in Table \ref{tb2}, confirms that both designs are computationally efficient due to the use of linear subsystem models despite nonlinear dynamics, ensuring the feasibility of online implementation.
Meanwhile, the distributed estimation scheme based on the linearized subsystem models is indeed more computationally efficient, which is consistent with expectations. The increase in the computation time stems from the higher-dimensional nature of the Koopman subsystem models in comparison to the linearized subsystem models, which, however, is at the expense of a significant reduction in the estimation accuracy.
}
  {\color{black}
\begin{rmk}
 The performance of the proposed method is dependent on the noise level in the dataset and the data sampling frequency. The established Koopman-based linear models can accurately capture the dynamic behaviors of the underlying nonlinear process when the noise remains within a reasonable range. In the meantime, the data frequency also can impact the performance of our data-driven method for both subsystem modeling and distributed state estimation of the proposed framework. The data needed for implementing the proposed subsystem modeling and distributed state estimation methods are expected to meet two criteria: 1) they should encompass informative transient trends of the dynamic process; 2) they should contain timely output measurements to ensure the real-time implementation of the proposed distributed moving horizon estimation approach, in accordance with the state estimate frequency requirement.
\end{rmk}}

\section*{Application to an agro-hydrological process}

In this subsection, the proposed methods on parallel subsystem identification and distributed MHE are applied to an agro-hydrological process through simulations for estimating the soil moisture at different locations of the soil profile. The modeling of this process and real-time estimation of the soil moisture are critical for informed closed-loop irrigation for sustainable water use and improved crop health.

\subsection*{Description of the process}
The agro-hydrological process incorporates the vadose zone of the soil. The dynamic interactions among the soil, crops, and the atmospheric environment within the hydrological cycle are considered. A schematic of this process is shown in Figure~\ref{fig:agro:schematic}. Water flow entering/leaving the soil is the known input to the process; this water flow can be caused by irrigation via sprinklers, rainfall, water extraction by the crop roots, evaporation, water runoff, and/or drainage.

Two standard assumptions are made for the agro-hydrological process: (1) soil properties, including color, texture, structure, porosity, and density, are homogeneous horizontally; (2) surface irrigation is conducted on the field uniformly.

\subsection*{First-principles model and discretization for data generation}
The aim is to leverage the proposed parallel subsystem model identification and distributed estimation method to the agro-hydrological process for data-driven subsystem modeling and distributed soil moisture estimation. In this subsection, we introduce a first-principles model based on the Richards equation \cite{richards1931capillary}; this high-fidelity model is only used for generating process data needed for implementing the proposed method.

To describe the vertical hydrological dynamical behaviors of the process, the Richards equation \cite{richards1931capillary} is introduced as follows:
\begin{equation}\label{eq:richards}
       \frac{\partial h}{\partial t} = \frac{1}{C(h)} \frac{\partial}{\partial z} \left[K(h)\left(\frac{\partial h}{\partial z}+1\right)\right]
\end{equation}
where $h$ (m) is the soil water pressure head; $z$ (m) is the vertical position inside the soil profile; $K$ (m/s) denotes the hydraulic conductivity which is dependent on $K$.
The dependence of $K$ on $h$ is characterized by the following equation \cite{mualem1976new}:
\begin{equation}\label{eq:Kh}
K(h)=
 K_{sat}S_e^\lambda\left[1-\left(1-S_e^{\frac{1}{m}}\right)^m\right]^2
\end{equation}
In (\ref{eq:Kh}), $m=1-1/n$, $\lambda$, $\alpha$ and $n$ are empirical parameters of the soil, $K_{sat}$ denotes the saturated hydraulic conductivity, and $S_e$ denotes the relative saturation which can be computed following:
\begin{equation*}
S_e = \frac{\theta(h) - \theta_r}{\theta_s-\theta_r}
\end{equation*}
where $\theta_s$ represents the saturated soil water content; $\theta_r$ represents the residual soil water content.
The relationship between soil moisture content $\theta(h)$ (m/m) and the pressure head can be characterized by the soil-water retention equation \cite{van1980closed}:
\begin{equation}\label{eq:retention}
\theta(h) =
     \left(\theta_{s}-\theta_{r}\right)\left(1+(-\alpha h)^n\right)^{-\left(1-\frac{1}{n}\right)}+\theta_{r}
\end{equation}
Further, the capillary capacity (i.e., $C(h)$ in Eq. (\ref{eq:richards})) can be calculated as \cite{van1980closed}:
\begin{equation}\label{eq:capacity}
    C(h) = n \alpha \left(\theta_{s}-\theta_{r}\right)\Big(1-\frac{1}{n}\Big)\left(-\alpha h\right)^{n-1}\left(1+\left(-\alpha h\right)^n\right)^{-\left(2-\frac{1}{n}\right)}
\end{equation}


The Richards equation in (\ref{eq:richards}) is a nonlinear partial differential equation (PDE) concerning both time and vertical coordinates. To obtain information on soil moisture at finite specific locations, the PDE-based model can be discretized. Specifically, finite difference is used to conduct time and space discretization to calculate numerical approximations of (\ref{eq:richards}). The soil moisture in each compartment remains homogeneous horizontally. This ensures the generated data can represent an accurate numerical approximation of the Richards equation in (\ref{eq:richards}). Following the discretization procedure and the boundary conditions adopted in Yin et al.\cite{yin2021consensus} and Bo et al. \cite{bo2020parameter}, a two-point forward difference can be conducted to approximate the time derivatives, and a two-point central difference can be conducted to approximate the spatial derivatives. With discretization in both time and space, the dynamic model in (\ref{eq:richards}) can be simulated to generate soil moisture samples for different locations of the soil profile. The generated samples will be used to identify the Koopman-based subsystem models in parallel.

\subsection*{Process setting and data generation}
The values of the parameters involved in (\ref{eq:richards}) are the same as those adopted in Carsel and Parrish \cite{carsel1988developing}.
A loam soil profile with a depth of $L=120$ cm is considered. In terms of space discretization, the entire soil profile is divided into 96 compartments as shown in Figure~\ref{figure:soil:profile}, in each of which the soil moisture is homogeneous. {\color{black}$\bigtriangleup z$ represents the size of the space step between successive nodes in discretization}. The capillary pressure head information in the discretized compartments constitutes a state vector of 96 state variables. Water flow enters the soil profile at the rate of $1.944\times 10^{-3}~ \text{m}/\text{hour}$ for eight hours every day, which is the same as the irrigation schedule in Yin et al. \cite{yin2021consensus} Free drainage takes place on the bottom boundary of the soil profile.

In the soil profile, there are 96 compartments of the same size after discretization, and historical data of the 96 states will be used for subsystem modeling. 
{\color{black}As depicted in the illustrative diagram of the soil profile in Figure~\ref{figure:soil:profile}, we consider that 16 permanent sensors are placed at the center points of the $(12\times i-10)$th and $(12 \times i)$th compartments, $i=1,2,\ldots,8$. Each sensor can provide real-time measurements of the soil moisture in the corresponding compartments.}
The 96 state variables of the process (i.e., the capillary pressure head in the 96 compartments) are divided into 8 subsystems, each of which is assigned 12 consecutive state variables. Specifically, $x_i = \big[x^{\left\{12\times i-11\right\}},\ldots, x^{\left\{12\times i\right\}} \big]^{\text{T}} $, $i=1,2,\ldots,8$, where $x^{\{j\}}$ is the $j$th variable of state vector $x$ counting from the top to the bottom. This way, each subsystem is assigned two physical sensors for measuring the capillary pressure head in the {\color{black}2nd and 12th} compartment of the same subsystem.

Unknown additive process disturbances and measurement noise following Gaussian distribution are generated, and then are made bounded and added to the discretized simulation model for generating data samples. We consider synchronous sampling of the soil moisture measurements every 1~min. Then, 9,600 data samples generated from simulating (\ref{eq:richards}) for an operating period of 160 hours are used for identifying the subsystem models based on the proposed method. In addition, 4,800 samples for an additional 80-hour operation are used for validating the model, and another 80-hour operating period is considered for illustrating the soil moisture estimation performance.

\subsection*{Subsystem modeling results}\label{section5:5}
Then, the proposed parallel subsystem modeling approach in proposed in the ``EDMD-based parallel Koopman subsystem identification method" subsection is applied to establish 8 subsystem models for the agro-hydrological process. State and input lifting functions $\phi^i$ and $\delta^i$ need to be specified for subsystem $i$, $i=1,2,\ldots,8$. As the states assigned to different subsystems have the same physical meaning, the lifting functions selected for different subsystems may be made identical. Inspired by the adopted basis functions in Narasingam and Kwon \cite{narasingam2019koopman}, the original subsystem state $x_i$, quadratic functions of $x_i$, and exponential functions of $x_i$ are chosen as the state lifting functions for each subsystem, that is,
\begin{equation*}
\phi^i(x_i) =\left[ {\begin{array}{*{20}{c}}
x_{i,1}&\ldots&x_{i,12}&x_{i,1}^2&\ldots&x_{i,12}^2&{e^{x_{i,1}}}&\ldots&{e^{x_{i,12}}}
\end{array}} \right]^{\text{T}},~\forall i=1,2,\ldots,8.
\end{equation*}
{\color{black}where $x_{i,j}$ denotes the $j$th element of the state of subsystem $i$, $\forall i=1,2,\ldots,8$, $\forall j = 1,2,\ldots,12$.}
The only known input to the process is the water flow entering/leaving the soil to the process. Since this input corresponds to the surface of the soil profile, this known input is assigned to the first subsystem.
Therefore, we only need to construct input lifting functions for the first subsystem, which are in the form of
$\delta^1(u_1) =\big[ {\begin{array}{*{20}{c}}
u_{1}&u_{1}^2&{e^{u_{1}}}
\end{array}} \big]^{\text{T}}$,
which accounts for the nonlinear dependence of the lifted subsystem state of the 1st subsystem on this known input.


As has been discussed, 9,600 training samples within a 160-hour operating period are used for identifying the Koopman-based subsystem models in parallel. Snapshot matrices ${\boldsymbol {x_i}}$, ${\boldsymbol {{\bar x}_i}}$, ${\boldsymbol {u_i}}$, and ${\boldsymbol {y_i}}$ of appropriate dimensions are created. Based on the selected state, input lifting functions, and the snapshot matrices, Algorithm~\ref{Algorithm:1} is implemented to identify the models for the 8 subsystems. It is noted that the snapshot matrices are constituted by scaled states and scaled manipulated inputs that are obtained following Steps~2.1 and 2.2 of Algorithm~\ref{Algorithm:1}. Eight Koopman subsystem models are constructed.
For each subsystem model in the form of (\ref{eq:general:subsysmodel:Linear:stochas}), the lifted state vector $z_i$ has 36 variables for $i=1,2,\ldots,8$; the lifted input vector $\tilde u_1$ for the 1st subsystem has 3 variables (since there is no known input to each subsystem $j$, $j=2,3,\ldots,8$).

Cross-validation of the established linear subsystem models based on 4,800 validation samples is performed. The trajectories of the actual soil moisture in selected compartments, and the corresponding open-loop predictions provided by the eight Koopman-based subsystem models are shown in Figures \ref{fig:Koopman_agro_fig4} and \ref{fig:Koopman_agro_fig5}, where the blue lines represent the actual soil pressure head in the selected compartments, and the red dashed lines are open-loop predictions of the soil pressure head information given by the identified Koopman subsystem models. Without correction/adaptation using actual soil moisture data, the linear subsystem models can collaboratively characterize the nonlinear dynamics of the agro-hydrological system accurately. Good predictions of the soil moisture in different compartments of the soil profile are obtained.
{\color{black}As shown in Figures \ref{fig:Koopman_agro_fig4} and \ref{fig:Koopman_agro_fig5}, the discrepancy between the actual states and open-loop predictions increases with the depth. The increase in the discrepancy arises due to the error propagation from the top compartment to the current compartment.}
It is worth mentioning that while we only present the estimation results for selected compartments of the soil profile, the Koopman-based subsystem models can provide accurate open-loop predictions of all the states (i.e., soil moisture in all the compartments of the soil profile).

\begin{rmk}
In general, there is no systematic procedure that can be followed to determine the state and input lifting functions.
Different lifting functions can indeed affect the accuracy of the resulting subsystem models. The selection of the lifting functions can be conducted through extensive trial-and-error tests, or neural networks that may be introduced and trained to account for the manually selected lifting functions in parametric forms; some relevant results were reported in Yeung et al. \cite{yeung2019learning} and Han et al.\cite{han2020deep}
\end{rmk}

\subsection*{Soil moisture estimation}

In this section, a partition-based distributed MHE scheme is developed based on the Koopman subsystem models, and the estimation results for the soil moisture in different compartments, especially the compartments where no permanent sensors are deployed, are presented.

A linear distributed MHE scheme consisting of 8 local estimators is developed. Each local estimator is designed following (\ref{eq:cost:function:problem1:1}).
The length of the estimation horizon for each local estimator is $N=4$. 
The weighting matrices are set as follows: $P_{i,0}=0.1\times{I}_{36}$, $Q_i = 0.01\times{I}_{36}$, and $R_i = 0.6\times{I}_6$, for $i=1,2,\ldots, 8$. Note that due to the scaling, the scaled subsystem states stay within the range $[0,1]$, which simplifies the selection of the tuning parameters of the MHE-based estimators.
The initial guesses of the soil moisture estimates provided to the local estimators are made different from the actual values of the respective initial subsystem states.
In the MHE-based local estimators, hard constraints are imposed on the soil moisture estimates provided by the MHE-based estimators for the subsystems, following the hard constraints considered in Yin et al. \cite{yin2021consensus} and Bo at al. \cite{bo2020parameter} {\color{black}Specifically, the estimates of the soil moisture are required to stay within $[-1.0, -1.0\times 10^{-6}]$ as has been employed in Yin et al. \cite{yin2021consensus}}. It is noted that these constraints are imposed on the first state to the twelfth state of each local estimator that represents the soil moisture in the 12 compartments of the same subsystem. No constraints need to be imposed on the 24 lifted state variables of the subsystems.

Next, we present the soil moisture estimation results for the entire agro-hydrological process provided by the distributed MHE. Figures~\ref{fig:estimate_DMHE_Argo:1} and \ref{fig:estimate_DMHE_Argo:2} show the trajectories of the actual soil moisture and the corresponding estimates in the selected compartments of the soil profile. The estimates can track trends of the actual soil moisture in the corresponding compartments, and the discrepancy between the actual values and the corresponding estimates remains small during the entire process operation.
{\color{black}
It is worth mentioning that despite the noticeable discrepancy between the actual states and open-loop predictions of the soil moisture in the 95th compartment (not measurable) shown in Figure \ref{fig:Koopman_agro_fig5}, the proposed distributed MHE method can still provide good state estimates. This is because the developed distributed MHE is a closed-loop estimation scheme that receives and takes advantage of the information of measurable states from the local subsystem and the interacting subsystems at each time instant, whereas the Koopman-based subsystem models provide open-loop predictions of the states without using any feedback information for correction/adaptation.}
The trajectory of the estimation error norm is presented in Figure~\ref{Error_irrigation}. The estimation error provided by the proposed method remains bounded during the operating period despite the use of data-driven linear subsystem models in the state estimator design.

{\color{black}
\begin{rmk}
The subsystem decomposition and configuration can affect the prediction performance of the Koopman-based subsystem models. In this work, in each of the two case studies, we consider that the entire process is decomposed by taking into account the physical topology of the process.
Meanwhile, in the existing literature, there have been systematic approaches to process decomposition and subsystem configuration for distributed estimation and control, including the community detection-based approaches with the aim of minimizing the strength of interactions among the configured subsystems \cite{yin2019subsystem,tang2018optimal}, the decomposition of the process into subsystems according to different time scales of the state variables \cite{yin2017distributed,wu2022economic}, and the hierarchical clustering \cite{kang2016control,heo2016control} which aim to identify the optimal subsystem designs to strengthen the connection between components within each subnetwork while weakening the connection between different subnetworks.  Some of these subsystem configuration approaches may be applied to decompose a large-scale process into smaller subsystems, as the basis of subsystem modeling using the proposed approach.

\end{rmk}}
\section*{Concluding Remarks}\label{Section:conclusion}
In this work, we addressed a data-driven distributed state estimation problem for general large-scale nonlinear processes that consist of interactive subsystems. The proposed framework comprises two key approaches, that is,  the parallel subsystem modeling approach and the distributed state estimation approach based on the identified subsystem models.
In parallel subsystem modeling, Koopman identification was leveraged as the basis of identifying linear subsystem models that interact and coordinate with each other to collaboratively approximate the nonlinear dynamics of the considered process. An implementation algorithm was proposed based on Extended Dynamic Mode Decomposition to construct the linear subsystem models in parallel. Both the states and the inputs of each subsystem are projected onto the higher-dimensional spaces. Linear subsystem models are constructed in higher-dimensional spaces. Then, we proposed a partition-based distributed state estimation approach. The established linear subsystem models are used to develop linear MHE-based state estimators, which communicate with each other and are evaluated in an iterative fashion, to provide real-time estimates of the states of the large-scale nonlinear processes. {\color{black}
Two case studies were conducted to illustrate the proposed framework. In the first case study, a benchmark chemical process example was used to demonstrate the effectiveness and superiority of the proposed approach. In the second case study, a large-scale agro-hydrolocal process was considered.} Eight linear Koopman subsystem models that interact with each other were established. The open-loop predictions given by the linear subsystem models track the actual state of the underlying nonlinear process accurately.
Based on the identified subsystem models and the proposed distributed estimation algorithm, a distributed soil moisture estimation scheme was formed. Accurate soil moisture estimates were achieved across all the compartments of the soil profile, without utilizing the first-principles process model in the estimation system design.

{\color{black}
It is noted when there is any available physical knowledge about a process, it will be favorable to incorporate the existing first-principles knowledge into Koopman modeling to build physics-enabled Koopman models, which constitutes one of potential future research directions.
}

\section*{Data Availability and Reproducibility Statement}

{\color{black}The numerical data used for generating Figures 3, 4, 7-11, and Table 2 is available in the Supplementary Material. The compressed file contains data related to the case studies on the chemical reaction process and the argo-hydrological process, including data used for the identification of Koopman subsystem models, maximum and minimum values of states utilized for scaling, the identified Koopman subsystem models, and the state estimates calculated based on the proposed method. The data can be used to reproduce and generate the figures and tables presented in this paper. Particularly, the RMSE in Table 2 can be obtained by utilizing the actual states , and the state estimates provided by the proposed distributed MHE based on the linearized subsystem models and the proposed Koopman subsystem models.
The subsystem matrices used for generating open-loop predictions for Figures 3, 7, and 8 can be derived by using the provided data and following Algorithm 1. The state estimates used to generate Figures 4, 9, and 10 can be obtained by implementing Algorithm 2.

\section*{Acknowledgment}

This research is supported by Ministry of Education, Singapore, under its Academic Research Fund Tier 1 (RG63/22), and Nanyang Technological University, Singapore (Start-Up Grant).

\renewcommand\refname{Literature Cited}

\newpage~
\begin{figure}[t]
\centerline{\includegraphics[width=0.76\textwidth]{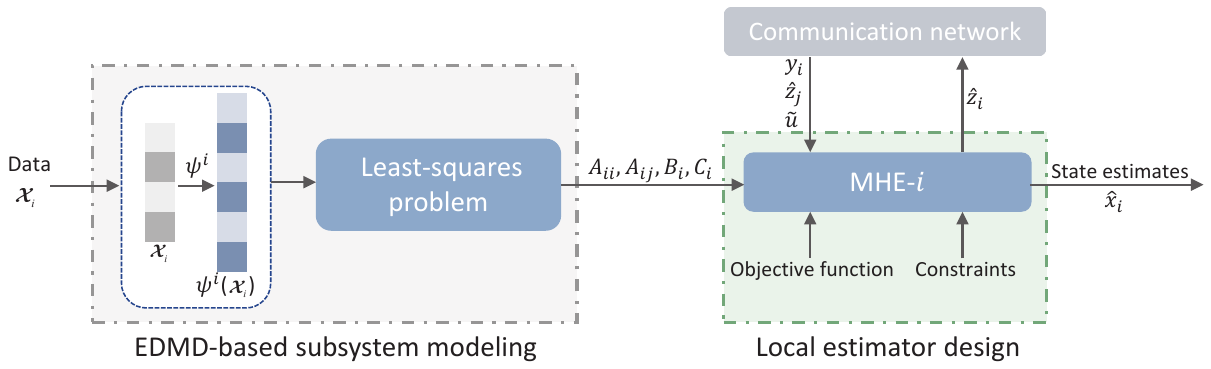}}
\caption{An illustrative diagram depicting the connection between the subsystem modeling and local estimator design of the proposed distributed estimation framework.}
\label{fig:diagram}
\end{figure}

\newpage~
\begin{figure}
    \centering
    \includegraphics[width=0.7\textwidth]{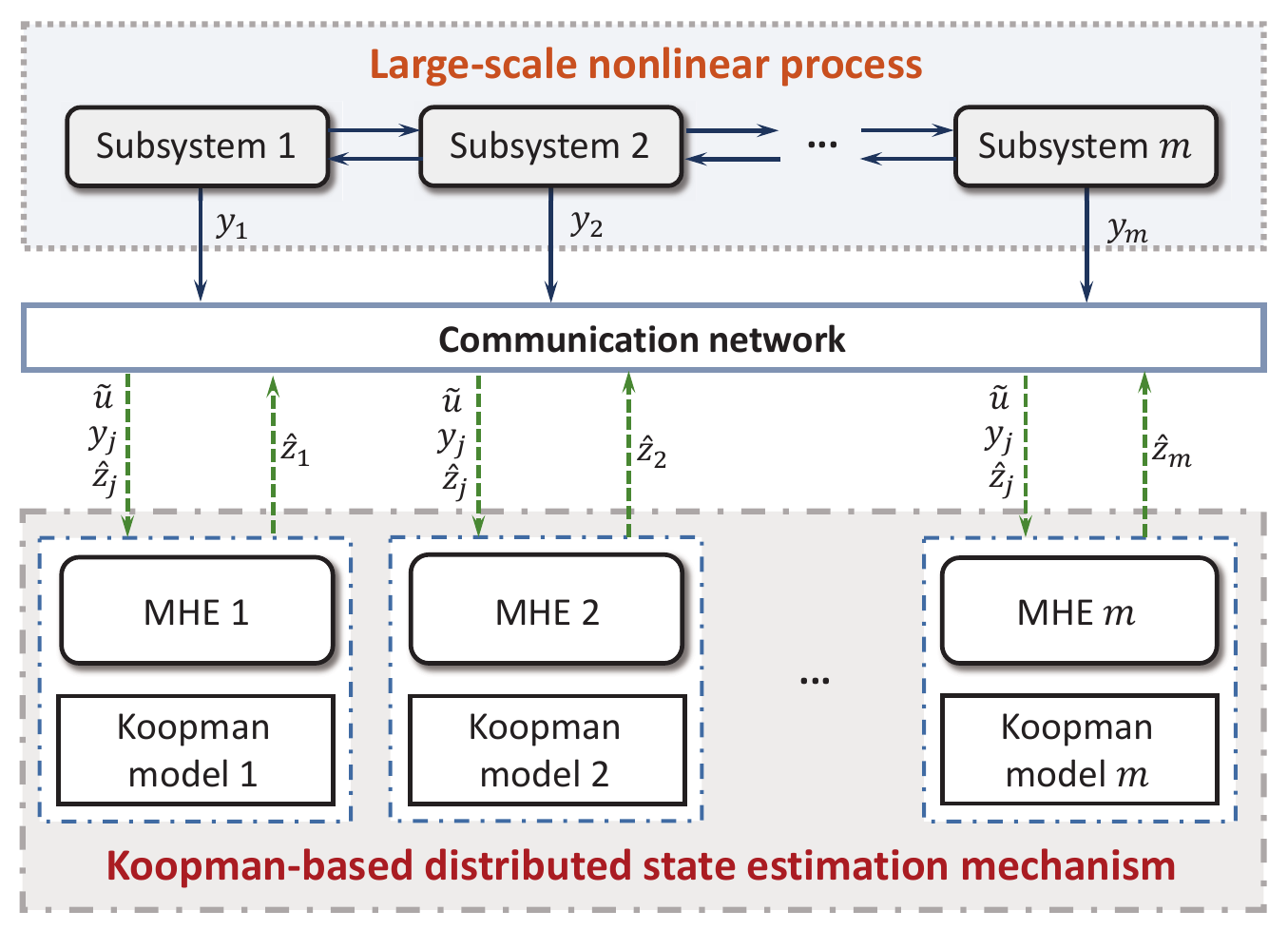}
    \caption{A schematic of the distributed state estimation scheme developed using Koopman subsystem models.}
    \label{fig:DMHE:schematic}
\end{figure}

\newpage~
\begin{figure}[tttt]
  \centering
  \includegraphics[width=0.75\textwidth]{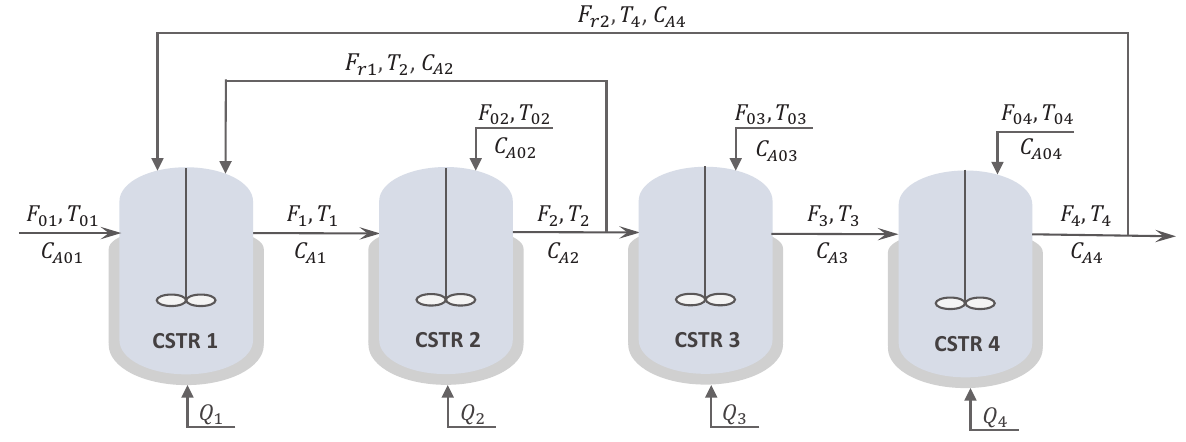}
  \caption{A schematic of the four-CSTR process.}\label{cstr}
\end{figure}

\newpage~
\begin{figure}[t]
\centerline{\includegraphics[width=0.75\textwidth]{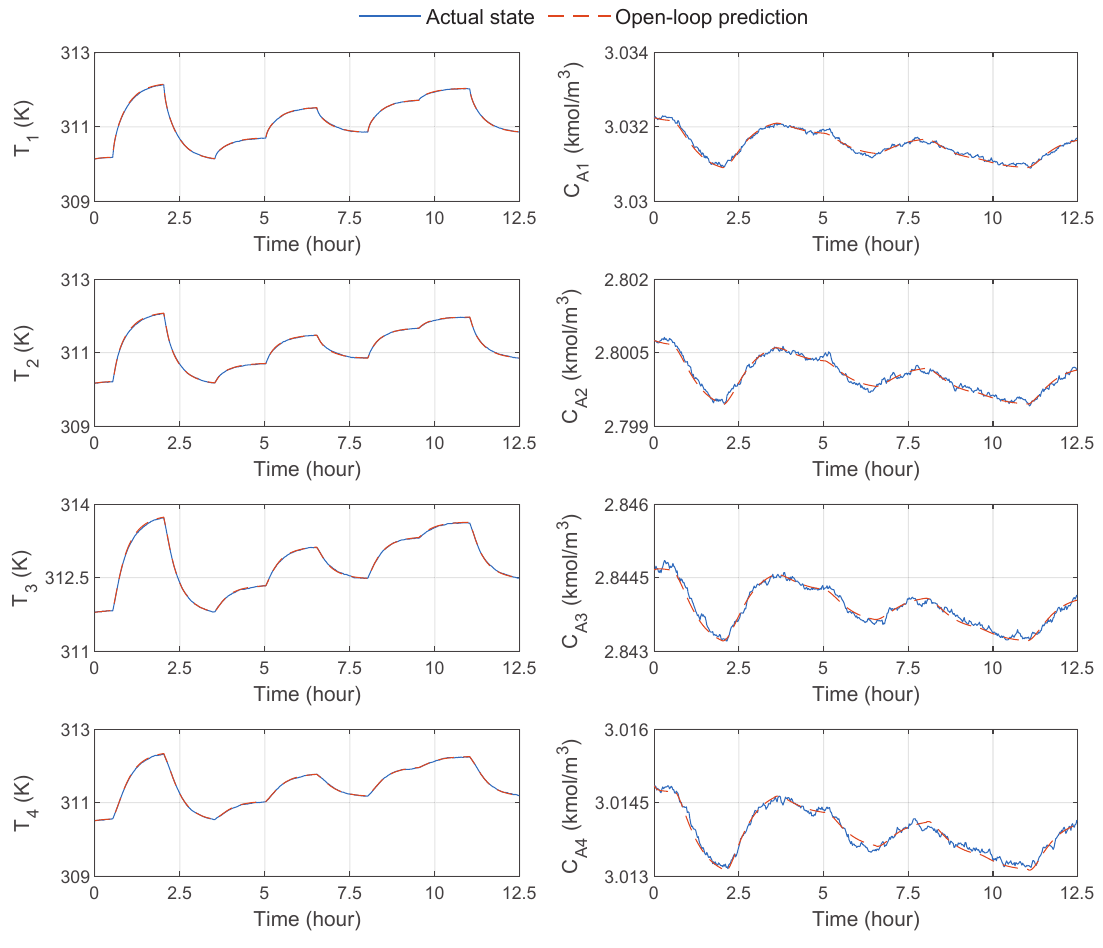}}
\caption{Cross-validation of the Koopman subsystem models for the four-CSTR process.}
\label{fig:cstr_validation}
\end{figure}

\newpage~
\begin{figure}[t]
\centerline{\includegraphics[width=0.75\textwidth]{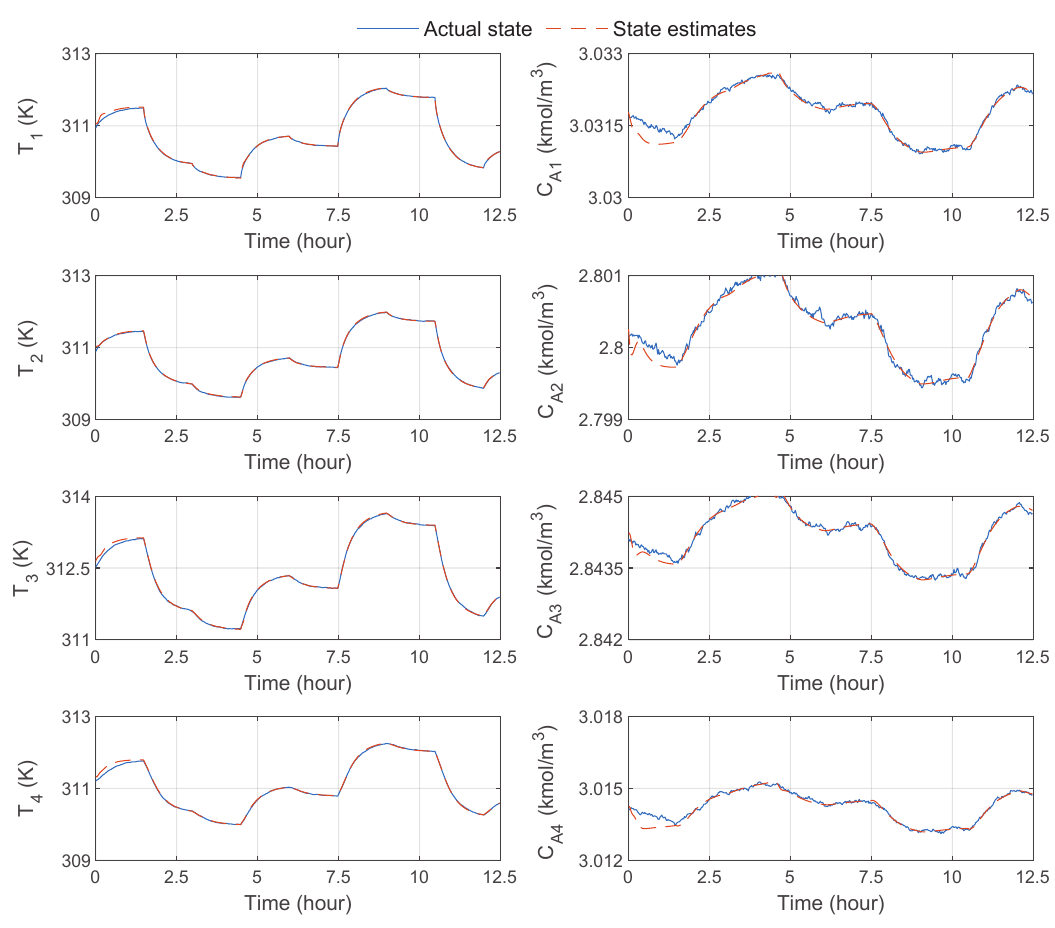}}
\caption{Trajectories of the actual system states and state estimates for four vessel CSTRs.}
\label{fig:cstr_estimation}
\end{figure}

\newpage~
\begin{figure}
    \centering
    \includegraphics[width=0.45\textwidth]{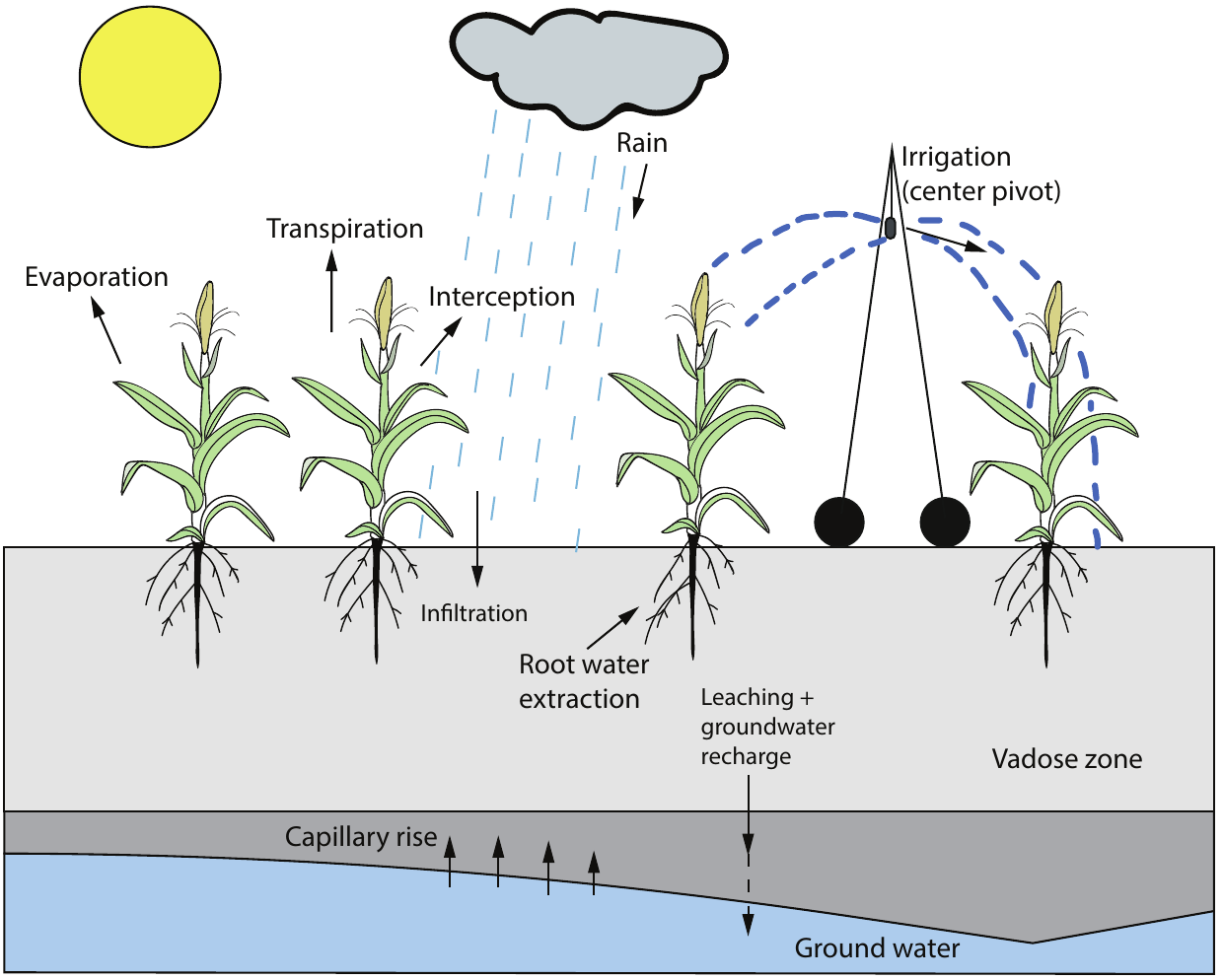}
    \caption{An illustrative schematic of the agro-hydrological process.}
    \label{fig:agro:schematic}
\end{figure}

\newpage~
\begin{figure}[t]
\centerline{\includegraphics[width=0.28\textwidth]{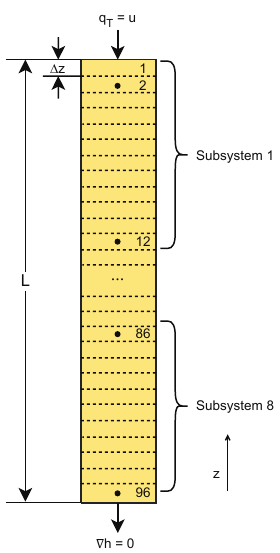}}
\caption{The compartments of the soil profile after vertical space discretization, and the locations of the sensors (black dots).}
\label{figure:soil:profile}
\end{figure}

\newpage~
\begin{figure}[t]
\centerline{\includegraphics[width=0.75\textwidth]{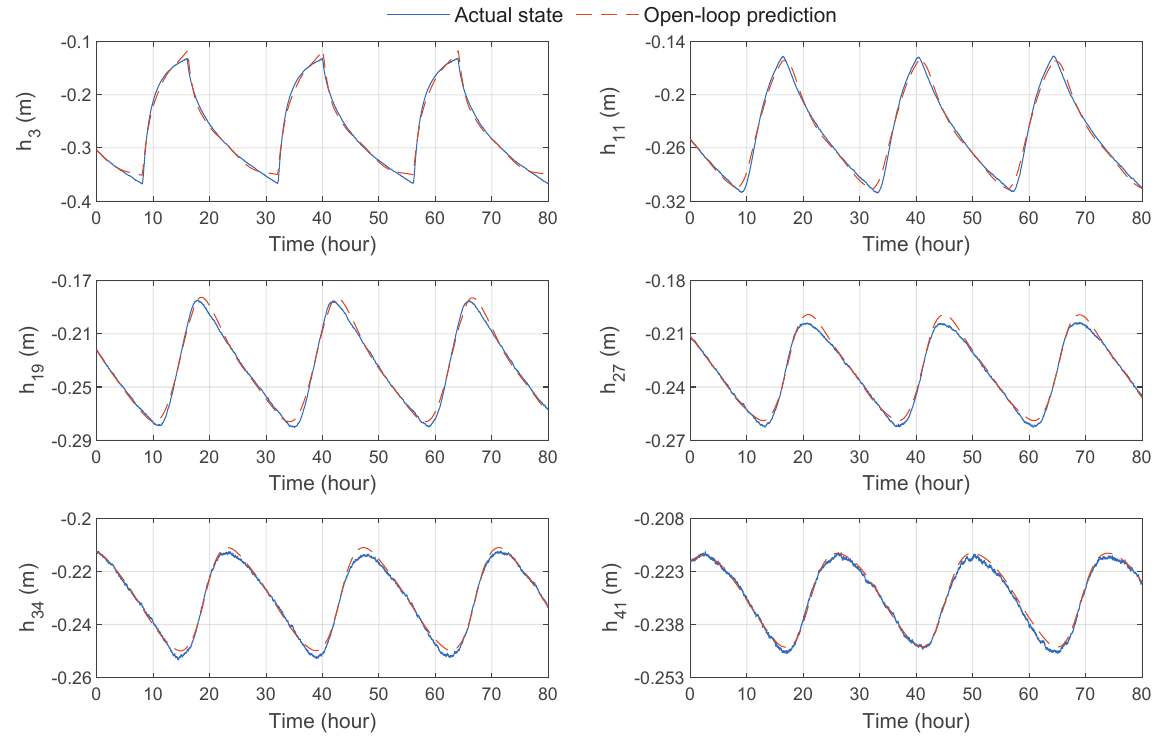}}
\caption{Cross-validation of the Koopman subsystem models (1st subsystem to 4th subsystem) for the argo-hydrological process.}
\label{fig:Koopman_agro_fig4}
\end{figure}

\newpage~
\begin{figure}[t]
\centerline{\includegraphics[width=0.75\textwidth]{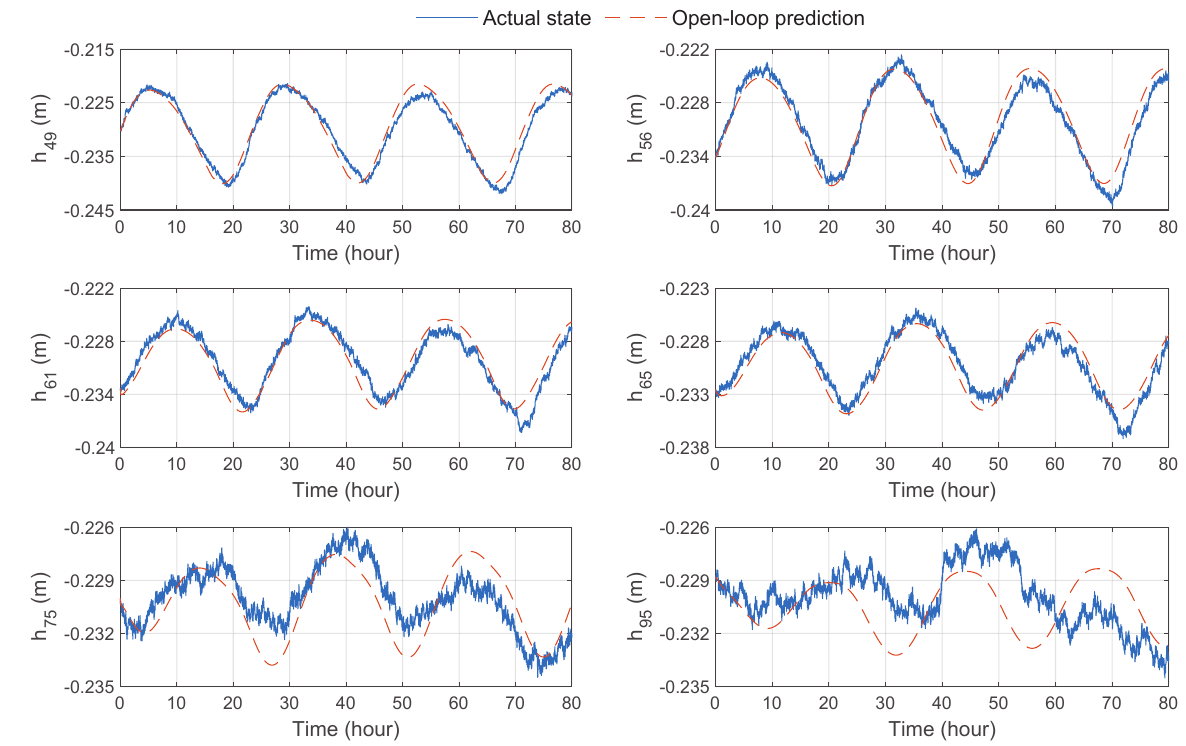}}
\caption{Cross-validation of the Koopman subsystem models (5th subsystem to 8th subsystem) for the argo-hydrological process.}
\label{fig:Koopman_agro_fig5}
\end{figure}

\newpage~
\begin{figure}[t]
\centerline{\includegraphics[width=0.75\textwidth]{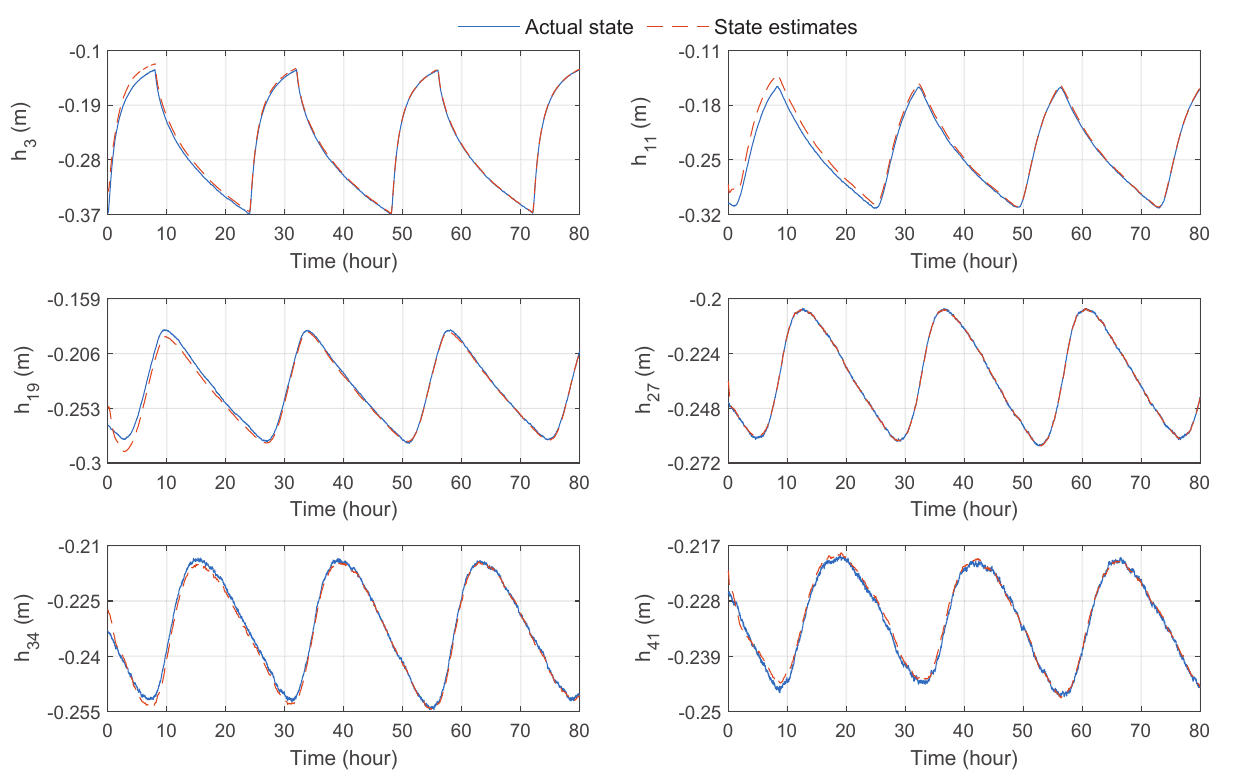}}
\caption{Trajectories of the actual system states and state estimates for the argo-hydrological process (1st subsystem to 4th subsystem).}
\label{fig:estimate_DMHE_Argo:1}
\end{figure}

\newpage~
\begin{figure}[t]
\centerline{\includegraphics[width=0.75\textwidth]{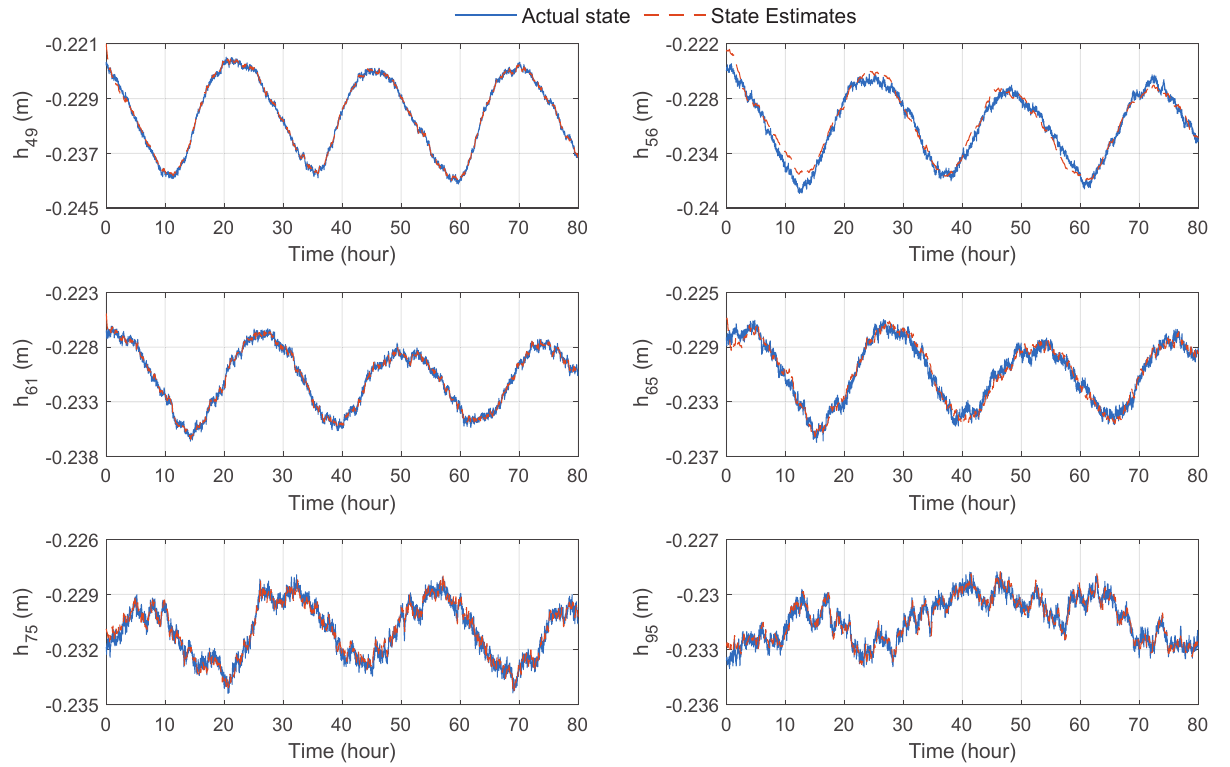}}
\caption{Trajectories of the actual system states and state estimates for the argo-hydrological process (5th subsystem to 8th subsystem).}
\label{fig:estimate_DMHE_Argo:2}
\end{figure}

\newpage~
\begin{figure}[t]
\centerline{\includegraphics[width=0.65\textwidth]{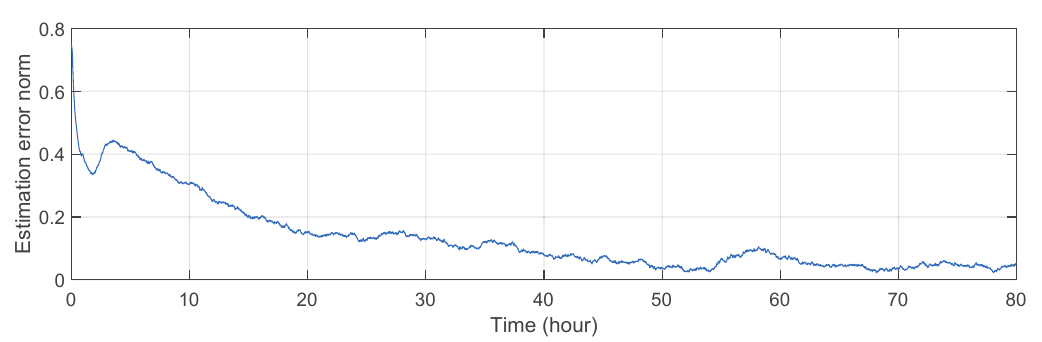}}
\caption{Trajectories of the estimation error norm for the argo-hydrological process.}
\label{Error_irrigation}
\end{figure}

\newpage~

\begin{table}
  \centering
    \caption{The initial state $x_{0}$ and the initial guess $\bar{x}_{0}$ for the process}\label{tbl:initial_states}
  \begin{tabular}{cccccccccc}
   \toprule[1pt]
   & $T_{1}$   & $C_{A1}$ & $T_{1}$  & $C_{A2}$ & $T_{3}$  & $C_{A3}$  & $T_{4}$  & $C_{A4}$ \\
   \midrule
   $x_{0}$ & 310.9387 & 3.0317 & 310.8962 & 2.8002 & 312.5167 & 2.8441 & 311.2047 & 3.0141 \\
    $\bar{x}_{0}$ & 311.0766 & 3.0318 & 311.1287 & 2.8003 & 312.7482  & 2.8440 & 311.5002 & 3.0139 \\
   \bottomrule[1pt]
\end{tabular}
\end{table}

\newpage~

\begin{table}
  \centering
    \caption{Comparison of computation time and estimation norm}\label{tb2}
  \begin{tabular}{c|c|cccccccc}
   \hline
  Estimation method & Computation time (s)   & RMSE \\
   \hline
   \makecell[c]{Distributed MHE \\based on Koopman subsystem models} & 0.1858  &  0.0135\\  
   \hline
   \makecell[c]{Distributed MHE \\based on linearized subsystem models} & 0.0653 & 1.5204 \\ 
   \hline
\end{tabular}
\end{table}

\end{document}